\documentclass[journal]{IEEEtran}
\usepackage{subcaption}
\usepackage[font=small,labelfont=bf]{caption}
\usepackage{amssymb}
\usepackage{comment}
\usepackage{amsmath}
\usepackage{nccmath}
\usepackage{amsfonts}
\usepackage{graphicx}
\usepackage{tabularx,booktabs}
\usepackage{soul}
\usepackage{float}
\usepackage{caption}
\usepackage{subcaption}
\usepackage{placeins}
\usepackage[hyphens,spaces,obeyspaces]{url}
\usepackage{hyperref}
\usepackage[sort&compress, numbers]{natbib}
\usepackage{lineno}
\usepackage[dvipsnames]{xcolor}
\usepackage{natbib}
\usepackage[margin= 0.7 in]{geometry}

\IEEEoverridecommandlockouts                              

\newcommand{\real}{\mathbb{R}}
\newcommand{\Path}{}
\newtheorem{theorem}{Theorem}
\newtheorem{lemma}[theorem]{Lemma}

\title{\LARGE \bf
A Linear Dynamical Perspective on Epidemiology: Interplay Between Early COVID-19 Outbreak and Human Mobility  
}
\author{Shakib Mustavee$^{1}$, Shaurya Agarwal$^{2}$, Chinwendu Enyioha$^{2}$, Suddhasattwa Das$^{3}$ 
\thanks{$^{1}$Shakib Mustavee is a Ph.D. student in the Department of Civil Environment and Construction Engineering of University of Central Florida, Orlando, USA
{\tt\small smustavee@knights.ucf.edu}} %

\thanks{$^{2}$Shaurya Agarwal and Chinwendu Enyioha are Assistant Professors at the University of Central Florida, Orlando, USA}
\thanks{$^{3}$ Suddhasattwa Das is a post doctoral fellow at George Masion University}

}

\begin{document}

\maketitle
\thispagestyle{plain}
\begin{abstract}
This paper investigates the impact of human activity and mobility (HAM) in the spreading dynamics of an epidemic. Specifically, it explores the interconnections between HAM and its effect on the early spread of the COVID-19 virus. During the early stages of the pandemic, effective reproduction numbers exhibited a high correlation with human mobility patterns, leading to a hypothesis that the HAM system can be studied as a coupled system with disease spread dynamics. This study applies the generalized Koopman framework with control inputs to determine the nonlinear disease spread dynamics and the input-output characteristics as a locally linear controlled dynamical system. The approach solely relies on the snapshots of spatiotemporal data and does not require any knowledge of the system’s physical laws. We exploit the Koopman operator framework by utilizing the Hankel Dynamic Mode Decomposition with Control (HDMDc) algorithm to obtain a linear disease spread model incorporating human mobility as a control input. The study demonstrated that the proposed methodology could capture the impact of local mobility on the early dynamics of the ongoing global pandemic. The obtained locally linear model can accurately forecast the number of new infections for various prediction windows ranging from two to four weeks. The study corroborates a leader-follower relationship between mobility and disease spread dynamics. In addition, the effect of delay embedding in the HDMDc algorithm is also investigated and reported. A case study was performed using COVID infection data from Florida, US, and HAM data extracted from \textit{Google community mobility data report.}
\end{abstract}

\section{INTRODUCTION}
Given how increasingly connected the world is, epidemics are becoming more of a commonplace. As we know and have come to see through the ongoing COVID-19 pandemic, the significant loss of lives, as well as the short and long-term economic impact, can be very devastating. Besides the loss of lives, the pandemic has also crippled global transportation, food supply, and challenged healthcare systems in ways not seen before. Understanding and forecasting the spread dynamics is a challenging task, in part because these are high-dimensional, nonlinear, and time-varying systems. In addition, the spreading process exhibits a multi-scale spatio-temporal phenomenon \cite{balcan2009multiscale,colizza2006role,quaranta2020understanding}. It depends on many exogenous variables, including human activity and mobility (HAM) and mitigation measures such as vaccination and face coverings adopted by people. HAM is considered a critical factor in the disease spread, given the fact that the effective reproduction number of the pandemic is highly correlated to mobility.

During the onset of the ongoing global pandemic, mitigation strategies revolved around imposing various restriction measures on human activity and mobility. Since the first `stay-at-home' order was issued in the United States on March 15, 2020, in Puerto Rico, similar executive orders issued by state and municipal authorities notably curbed travel demand and thus potentially limiting the community spread of COVID-19. Similarly, governmental agencies worldwide have imposed lockdown and introduced various social isolation strategies for controlling the spread of coronavirus. The underlying rationale for these restriction strategies, such as lockdown and social isolation, is to reduce the scope of direct interpersonal contacts, which on the other hand, adversely impact human activity and mobility, in turn, slows down the disease transmission rate. Thus dynamics of COVID-19 spread and the dynamics of HAM are intertwined via an intricate relationship.

Despite the close connections between the spread of a pandemic and mobility, obtaining a quantifiable relationship between them is challenging because the spread dynamics of a pandemic such as COVID-19 depend on various other factors such as social distancing mask-wearing, mutation of the virus, etc. Moreover, mobility is a multi-modal service, which means each mode of mobility has a different mechanism to impact the disease spread dynamics. Thus, Linka et al. characterized mobility as a `global barometer’ of COVID-19 \cite{linka2020global}.

\begin{figure}[!htbp]
\centerline{\includegraphics[width=0.45\textwidth]{\Path 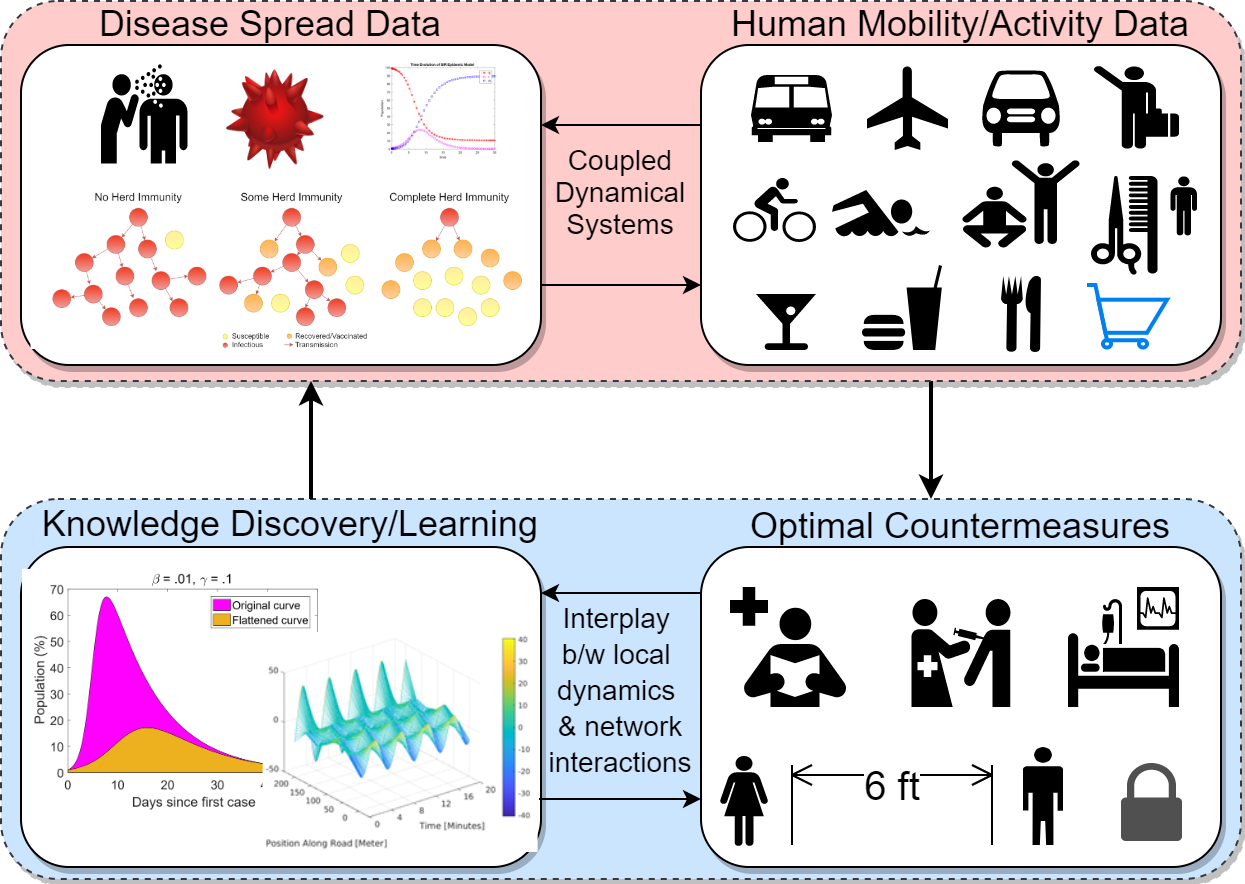}}
\caption{Data-driven discovery of coupled system}
\label{fig:coupled}
\end{figure} 

Although the literature has, for many decades, proposed ways of modeling epidemic processes in human populations, several factors that affect the rate of spread or control are not often accounted for. For example, human mobility and activity levels in an outbreak can affect how fast the outbreak is contained or how quickly it spreads. While this knowledge, in part, informs the preliminary soft measures typically taken by health policymakers such as social distancing and lockdown, which prevents mobility or travel of individuals between infected and uninfected areas, the degree to which such external factors affect the pace of spread are not often wholly known. A significant research gap that arises here is establishing the interplay between the pandemic dynamics and the HAM changes due to adopted countermeasures (lockdown, social distancing, etc.). This knowledge gap makes the existing control strategies deviate from arguably optimal approaches based on the actual nonlinear dynamics. For a highly infectious disease like COVID-19, a paradigm shift in characterizing the spreading dynamics is necessary. To contain a resurgence of the outbreak using scarce or limited resources (such as vaccines and ventilators),  a reliable approach of integrating HAM into the spread models aided by novel data-driven tools within a rigorous mathematical framework is necessary. The output of the proposed tightly integrated, the equation-free approach can provide more robust analysis and serve as a helpful tool for policymakers.  

\begin{figure}[!htbp]
\centerline{\includegraphics[width=0.45\textwidth]{\Path 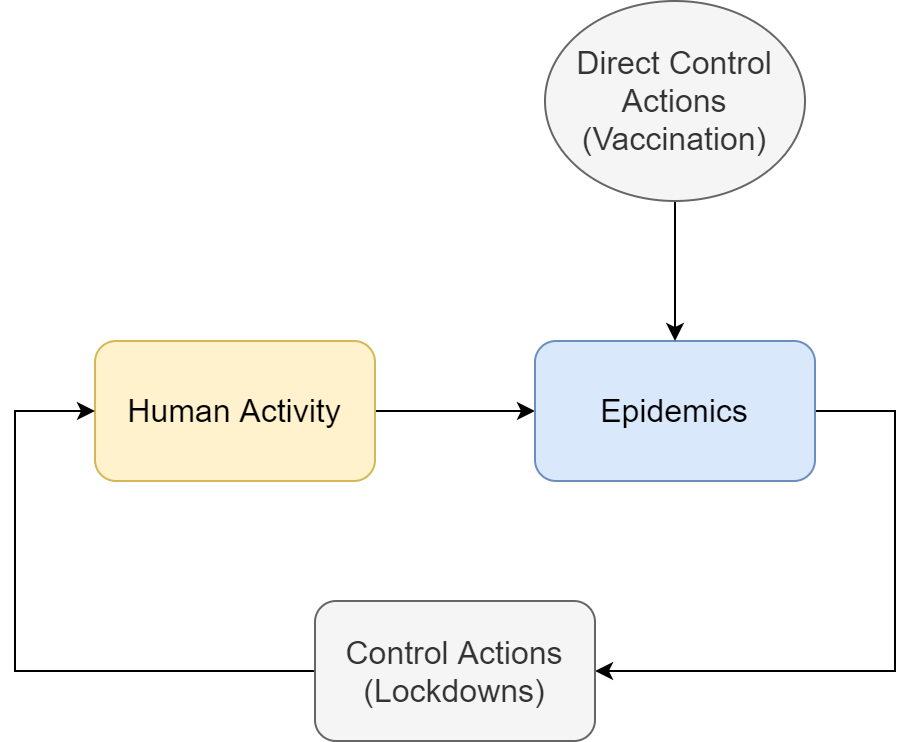}}
\caption{Feedback loop arrangement for disease system}
\label{fig:input-output}
\end{figure}

In this paper, We uncover the interconnections between the simultaneous evolution of two systems -- HAM and disease spread -- by treating them as a coupled dynamical system (see Figure \ref{fig:coupled}). This research builds a system discovery framework through the Koopman operator framework exploiting Hankel dynamic mode decomposition with control (HDMDc) algorithm for understanding the dynamics of the COVID-19 disease and its interplay with HAM. Figure~\ref{fig:input-output} illustrates the arrangement of the two systems in a feedback loop set. HAM and epidemics spread system are coupled in a cascaded fashion, where the current state of epidemics (e.g. number of daily infections) triggers a feedback control action (e.g. lockdowns), which affects the HAM system. This feedback through the HAM system feeds back into the epidemics system. Whereas other inputs (e.g. vaccinations) may affect the epidemics system directly.

As a case study, this paper studies the early dynamics of COVID-19 in the state of Florida, US, and develops a locally linear model using large-scale snapshots of spatio-temporal infection data and  HAM data extracted form \textit{Google community mobility data report}. Integrating HAM data into the framework will highlight the undercurrents of the disease evolution as the economy reopened.



\textit{Contributions:} Contributions of this paper are summarized as follows:
\begin{itemize}
     
\item We formulate and study human activity, mobility, and epidemics spread as cascaded coupled dynamical systems.
\item The proposed method incorporates the impact of HAM into infectious disease dynamics.
\item The developed method for system identification yields a locally linear equation-free model.
\item A novel technique is proposed to quantify mobility as a representation of human activity, which is then used as a control input.
\item We also identify the effect of delay embedding on the model accuracy and prediction results.
\end{itemize}

\textit{Outline:} Rest of the paper is organized as follows: Section~\ref{literature review}  discusses the background of the problem and the current state of the art, Section~\ref{sec:math} presents the adopted mathematical framework, and Section~\ref{data source} discusses the characteristics of the data and its preparation details. Finally, in Section-\ref{results} we will illustrate and explain the results of our analysis.

\section{Background and Literature Review}\label{literature review}

This section briefly reviews some relevant literature and approaches on mathematical models in epidemiology (\ref{Modeling}), explores the connections between mobility and epidemic spreading (\ref{Mobility and epidemiology}), and highlights the application of DMD-type algorithms in various random dynamical system identification including epidemiology and transportation (\ref{DMDc}).         

\subsection{Mathematical Models in Epidemiology and COVID-19} \label{Modeling}

Although one of the earliest applications of mathematics in epidemiology is traced back to the mid of $18^{th}$ century demonstrated by Daniel Bernoulli \cite{bernoulli1760essai}, deterministic epidemiology modeling came into being at the beginning of the $20^{th}$ century \cite{hethcote2000mathematics}; however, these models were not built by mathematicians but by public health physicians who laid the groundwork of developing compartmental models between $1900$ and $1935$ \cite{brauer2019mathematical}. From the middle of the twentieth-century application of mathematics in epidemiology has been significantly increased.  

Since the onset of the global pandemic in December 2019, various mathematical models and techniques have been applied to understand, explain and predict its spreading dynamics. Rahimi and \textit{et al.} have presented an review on COVID-19 prediction models in \cite{rahimi2021review}. The proliferation of models at the same time has raised a lot of criticisms and doubts, as they often speculate contradictory prediction results. In \cite{james2021use} authors discussed key limitations of mathematical models used in interpreting epidemiological data and public decision making. This research presented several validation approaches to corroborate the speculations of these analyses.     

Mathematical methods used to model COVID-19 generally fall into two categories: (i) statistical models based on machine learning and regression (ii) mechanistic models such as SIR (susceptible-infected-recovered) or SEIR (susceptible-exposed-infected-recovered) type models and their various derivatives \cite{holmdahl2020wrong}. A comparative study between the two classes of models is presented in \cite{kuhl2020data}. While machine learning models, which fall under statistical models, are black-box approaches that rely solely on a significant amount of data and incorporate no inherent feature of the disease, mechanistic models exploit interacting disease mechanisms and incorporate disease-specific information. Due to the inherently different nature of the approaches, respective application scopes are also different. For example, statistical models are effective in short-term predictions, while mechanistic models are more suitable for long-term prediction horizons. The application of SEIR-type models in analyzing and predicting the COVID-19 outbreak has been widespread. Some examples can be found be in \cite{peirlinck2020outbreak,he2020seir,cobey2020modeling}.

\subsection{Mobility and Epidemiology}\label{Mobility and epidemiology}
The relationship between mobility and the spread of infectious disease has been well established \cite{bajardi2011human,khan2009spread, cauchemez2011role,herrera2011multiple,espinoza2020mobility}. This was a subject of paramount importance even in the pre-COVID-19 era. Mobility restriction strategies like cordon sanitaire were implemented for controlling various epidemics such as bubonic plague (1666) \cite{race1995some}, yellow fever (1793, 1821, 1882 \cite{kohn2007encyclopedia}, and cholera (1830, 1884) \cite{cetron2005public}. Some of the early efforts that estimated the impact of mobility through mathematical analysis on disease outbreak include \cite{baroyan1971computer,rvachev1985mathematical}, which are followed by few other studies. The relationship between mobility and disease spread are reported in \cite{bajardi2011human,khan2009spread,cauchemez2011role,herrera2011multiple}. In \cite{espinoza2020mobility} 
authors illustrated the relationship between mobility restriction and epidemic at various scales and levels.   

One of the earliest attempts to correlated mobility trends with COVID-19 transmission was reported in \cite{xiong2020mobile}. This research showed that there lies a positive relationship between mobility inflow and the number of COVID-19 cases. A prediction model of COVID-19 using community mobility reports was proposed in \cite{wang2020using}. It used a partial differential equation model and integrated Google community mobility data with it. The model captures the \textit{combined effects of transboundary spread among county clusters in Arizona and human actives}. In \cite{zengspatial} the authors applied the Poisson time series model to explore the connection between population mobility and COVID-19 daily cases. The model was simultaneously applied at the county level and state level in South Carolina. On the other hand, the effect of \textit{human mobility trend under non-Pharmaceutical interventions} was investigated in \cite{hubig}, where a generalized additive mixed model (GAMM) was proposed. In \cite{iacus2020human} it was shown that the spread dynamics of COVID-19 during the early stage of the outbreak is highly correlated to human mobility. In this research, mobility data was obtained from mobile data.  

By using the susceptible-exposed-infected-recovered (SEIR) model, the impact of air traffic and car mobility on COVID-19 dynamics in Europe was evaluated by Linka et al. \cite{linka2020global}. They used a standard SEIR compartment model with a network structure for capturing local dynamics. To integrate the effect of mobility in the model, they applied a hyperbolic tangent-type ansatz. The authors argued that local mobility has a high correlation with the reproduction number. Also, they showed that \textit{mobility and reproduction are correlated during the early stages of the outbreak but become uncorrelated during later stages}. Hence, they advocated that local mobility can be used as a quantitative metric for prediction and identification stages. Another insightful study was conducted in \cite{linka2020outbreak}, where the authors presented their arguments with quantitative analysis on travel restriction in controlling an outbreak. They also applied the SEIR model and integrated mobility into the model. Linka et al. have found a lead-lag relationship between mobility and the spread of COVID-19. Similar results were obtained in \cite{kuhl2020data}.

\begin{figure}[!htbp]
\centerline{\includegraphics[width=0.45\textwidth]{\Path 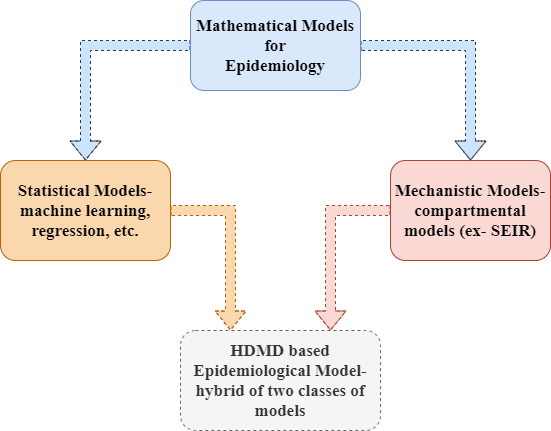}}
\caption{Classification of epidemiological models and their connection with the proposed HDMDc based model}
\label{fig:epidemics_model}
\end{figure}

\subsection{Application of DMDc in System Identification} \label{DMDc}

The application of dynamic mode decomposition (DMD) type models in epidemiology was first demonstrated in \cite{proctor2016dynamic}. In this research, the DMDc was applied to historical data to model the spread of infectious diseases like flu, measles, and polio and explore its connection with the vaccination. This study demonstrated that exogenous inputs of dynamics such as vaccination in case of epidemic spread could be incorporated with the original dynamics by using DMDc. 

The application of Hankel matrix structure in system identification was demonstrated in \cite{fazel2013hankel}. Like disease spread, traffic flow is also a dynamical system that is highly nonlinear and prone to randomness. In \cite{avila2020data} the authors extracted spatio-temporal coherent structures of highway traffic dynamics by using the concept of so-called Koopman modes. They applied Hankel DMD for computing Koopman modes and used the estimated modes for traffic dynamics forecast. Besides, this technique can be applied to capture various traffic dynamics of signalized intersections. In \cite{ling2018koopman} HDMDc was used to identify the underlying dynamical system of traffic queue. Furthermore, they made predictions on traffic queue length and applied Koopman spectral analysis to determine queue length instability. Another interesting application of DMD in random dynamical system identification is demonstrated in \cite{boskic2020koopman} where it was applied in infrastructure energy assessment. In this work, the authors used HDMD to extract insights into the thermal dynamics of an airport.

Figure \ref{fig:epidemics_model} illustrates the relationship between the major modeling paradigms used in epidemiology and their connection with the proposed approach. The advantages of DMD-type algorithms are summarized as follows:
\begin{itemize}
    \item DMD-type algorithms yield a locally linear model.
    \item DMD-type algorithms are purely data-driven and require no physical knowledge about the system.
    \item They do not require parameter estimation.
    \item They provide an equation-free approach.
    \item These approaches can be modified to incorporate input-output characteristics of the system.
\end{itemize}

\section{Mathematical Framework} \label{sec:math}
This section will present the mathematical framework of the proposed methodology, problem formulation, discussion on DMD type techniques such as HDMD and HDMDc, and the issue of delay coordinate embedding.

\subsection{Problem Formulation}\label{sec:data_driven}

To interpret and analyze the data that we are dealing with, we make the assumption that the data is generated by an underlying dynamical system. We assume that there is a phase space $\Omega$ and a map $F:\Omega\to \Omega$.  The transformation $F$ generates a dynamics on the phase space $\Omega$. Given any initial point $\omega_0\in \Omega$, there is a trajectory 
\begin{equation} \label{dynamics}
\left\{ \omega_{i+1} := F^i \omega_i \;:\; i=0,1,2,\ldots \right\} .
\end{equation}
We only assume the existence of $(\Omega, F)$, but not their explicit form. The information about $\Omega$ will obtained through a measurement / function $\mathcal{X} : \Omega \to \mathbb{R}^n$. This measurement generates the sequence of data points in $\mathbb{R}^k$
\[ \left\{ x_i := \mathcal{X}(\omega_i)\;:\; i=0,1,2,\ldots  \right\} \]
These assumptions make up the \emph{data-driven} framework, it enables a parameter-free reconstruction of the dynamical systems. Note that the measurement $\mathcal{X}$ is not necessarily one-to-one, thus the data-points $x_i$ may not correspond to a unique underlying state in $\Omega$. In particular, it may not be possible to connect the $x_i$ with a dynamical rule of the form $x_{i+1} = \tilde{F}(x_i)$. $x_i$ should be interpreted as a partial observation of the true state in $\Omega$, in the $i$-th time frame.

\subsection{Delay-coordinate embedding}
The method of delay coordinates incorporates a number $h$ of time shifted versions of the map $\mathcal{X}$ to obtain a higher dimensional map
\[ \mathcal{X}^{(h)} : \Omega \to \real^{k(h+1)},\] \[\mathcal{X}^{(h)}(\omega) := \left( \mathcal{X}(\omega), \mathcal{X}( F^1 \omega), \ldots, \mathcal{X}(F^h\omega) \right) . \]
Thus the delay coordinated version of each point $x_n$ is
\[ x_n \leftrightarrow x_N^{(h)} := \left( x_n, x_{n+1}, \ldots, x_{n+h} \right) . \]
The main purpose of delay coordinates is that they produce an \emph{embedding} of the dynamics, as shown in the lemma below.

\begin{lemma}[Delay-coordinate embedding \cite{sauer1991embedology}]  \label{lem:delay_coord}
Given a diffeomorphism $F:\Omega \to \Omega$, for almost every function $\mathcal{X}:\Omega\to \real^n$, there is an integer $h_0$ such that for every $h>h_0$, the function $\mathcal{X}^{(h)} : \Omega \to \real^{n(h+1)}$ obtained by incorporating $h$ delay coordinates into $\mathcal{X}$, is an embedding / one-to-one map of $\Omega$ into $\real^{n(h+1)}$.
\end{lemma}

This statement above is for ``almost every" measurement functions $\mathcal{X}$, meaning that it holds for typical $\mathcal{X}$ \cite{hunt1992prevalence} 
For such $\mathcal{X}$, the statement says that there is some choice of $h$ by which every value of $\mathcal{X}^{(h)}$ is the unique representative of an underlying state of $\Omega$. The image of $\mathcal{X}^{(h)}$ is thus an embedding or one-to-one image of $\Omega$ in the Euclidean space $\real^{n(h+1)}$. The size of $h$ would depend on $\mathcal{X}$ and $\Omega$.

\subsection{Dynamic mode decomposition}\label{sec:DMD}
Dynamic mode decomposition (DMD) is a tool for data-driven discovery of dynamical systems. One way to look at it is as a union of spatial dimensionality reduction and Fourier transformation in time. Let $X^{(h)}$ be the $n(1+h) \times m$ matrix formed by collecting $m$ consecutive snapshots $x_i^{(h)}$ :
\begin{equation*}
 X=
\begin{bmatrix}
\mid & \mid & & \mid\\
x_{1}^{(h)} & x_{2}^{(h)} & ... & x_{m}^{(h)} \\
\mid & \mid & & \mid\\
\end{bmatrix}
\end{equation*}

Similarly, let $X'$ be $n(1+h) \times m$ time shifted version of $X$. 
\begin{equation*}
X'=\begin{bmatrix}
\mid & \mid & & \mid\\
x_{2}^{(h)} & x_{3}^{(h)} & ... & x_{m+1}^{(h)}\\
\mid & \mid & & \mid\\
\end{bmatrix} .
\end{equation*}
%
The method of DMD takes a locally linear approximation of the dynamics \eqref{dynamics} by assuming a linear relation of the form 
\begin{equation}
 X' = AX, \label{Linear Model}
\end{equation}
where the matrix $A$ represents the discrete-time locally linear dynamics. It is a best fitting operator, which minimizes Frobenius norm of equation \eqref{Linear Model}. Suppose $X$ has the SVD (singular value decomposition) $X=U \Sigma V^*$. Then the Moore Penrose inverse of $A$ can be computed as
\begin{equation} \label{Pseudo}
A = X' ({U \Sigma V^*})^{-1} = X'V \Sigma^{-1} U^*
\end{equation}
DMD is thus a regression algorithm \cite{erichson2019compressed} and it produces locally linear approximation of the dynamics. However, in a few instances DMD fails to capture the dynamical nature of the system, e.g., in case of a standing wave. Hankel DMD (HDMD) is a variant of DMD applied on a time-delayed Hankel matrix which increases the order of the underlying dynamical system, thereby aiding in the estimation of hidden oscillatory modes.

\subsection{Dynamic mode decomposition with control} \label{sec:HDMDC}
DMD with control (DMDc) is a modified version of DMD which takes into account  both the system measurements and the exogenous control input to uncover the input-output characteristics and the underlying dynamics. Hankel DMD with control (HDMDc) is defined in a similar fashion and is the application of DMDc algorithm on the time delayed coordinates. If there is an external control input $u_k \in \real^q$ acting on the system, then the autonomous system \eqref{dynamics} becomes
\[ \omega_{k+1} = F_c( \omega_k, u_k ), k=0,1,2,3,\ldots  \]
Our linear model involving the measurement maps becomes 
\begin{equation}\label{system_A_B}
{x}_{k+1}^{(h)}  = A x_k^{(h)} +B u_k .
\end{equation}
If the $u_k$ are collected into an $s\times m$ matrix $\Upsilon$, we get the system of linear equations
\begin{equation} \label{DMD_Control}
    X' = A X +B \Upsilon
\end{equation}
%

\subsection{Connections with Koopman operator} \label{sec:Koop}
Although the original dynamics is by no means linear, \eqref{system_A_B} represents a finite dimensional approximation of a linear formulation of the dynamics, called the \emph{Koopman formulation}. The Koopman operator $\Psi$ is essentially a time-shift operator. It operates on functions instead of points on the phase space $\Omega$. Given any function $f:\Omega \to \real$, $\Psi f$ is another function defined as
\begin{equation} \label{eqn:def:Koop}
(\Psi f)(z) := f\left( F z\right) , \quad \forall z\in\Omega ,
\end{equation}
where $F$ is the underlying dynamical system \eqref{dynamics}. The Koopman operator converts any nonlinear dynamical system into a linear map. Thus all the tools from operator theory / functional analysis can be brought into the study of dynamics \cite{DasGiannakis_delay_2019,DGJ_compactV_2018,das2020koopman}. However, the original dynamics which is usually on a finite dimensional phase-space is converted into dynamics on some infinite dimensional vector space. 
The properties of $\Psi$ depends on the choice of vector/function space. Some common choices of function spaces are $C(\Omega)$ the space of continuous functions, or $C^r(\Omega)$, the space of $r$-times differentiable functions on $\Omega$. We shall use $L^2(\mu)$, the space of square-integrable functions with respect to the invariant measure $\mu$ of the dynamics. 
Now note that the columns of matrix $X$ are repeated iterations of $\Psi$ :
\[ X = 
\left[ (\Psi^0 \mathcal{X}^{(h)}) ( \omega_0), (\Psi^1 \mathcal{X}^{(h)}) ( \omega_0), \ldots, (\Psi^{m-1} \mathcal{X}^{(h)}) ( \omega_0) \right] , \]
for some unknown initial state $\omega_0\in \Omega$. Similarly
\[ \Upsilon = \left[ (\Psi^0 u) ( \omega_0), (\Psi^1 u) ( \omega_0), \ldots, (\Psi^{m-1} u) ( \omega_0) \right] , \]
for some unknown control-measurement $u : \Omega \to \real^q$. Thus both $A$, $B$ record the action of $U$ on finite dimensional Krylov subspaces. This justifies \eqref{system_A_B} as a finite rank approximation of the infinite-dimensional operator $\Psi$ and thus of $F$. If we set $\Omega = \begin{bmatrix}X & \Upsilon\end{bmatrix}^\top$ and $G = \begin{bmatrix} A & B \end{bmatrix}$, the equation becomes
\begin{equation} \label{augmented}
    X' = G\Omega
\end{equation}
To solve this, we reuse notation to take the SVD $\Omega = U \Sigma V^*$ and then set  
\begin{equation} \begin{split} \label{eqn:sdn38}
A &:= X'V \Sigma^{-1} X^* \\
B &= X'V \Sigma^{-1} \Upsilon
\end{split}\end{equation}
In practice, if the size of the data is limited to $N$ time samples, then if we choose $h$ delay coordinates, then $m$ can be at most $N-h$.

\section{Data Sources and Preparation}\label{data source}

This section discusses data sources and data preparation methods used in this research. It is divided into two subsections. \ref{early_dev} provides a timeline for state government response and case counts, \ref{data_des} provides the details of the data used in the case study, while \ref{Control Inputs} introduces the proposed technique of generating control inputs from the mobility data. 

\subsection{Early Developments in Florida} \label{early_dev}
After the first reported cases on March 1, 2020, the state of Florida entered into a public health emergency followed by a series of restrictions and shutdowns on bars and recreation facilities from March 17, 2020, to March 31, 2020. On April 1, 2020, the Governor of Florida issued a stay-at-home order. In the following two weeks number of reported cases gradually decreased, and subsequently, beach saloons, bars, and educational institutes gradually received permission to reopen from April 17, 2020, to June 10, 2020. However, as the number of cases rapidly increased, the state government shut down all bars on June 26, 2020.

\begin{figure*}[!htbp]
\centerline{\includegraphics[width=1.2\textwidth]{\Path 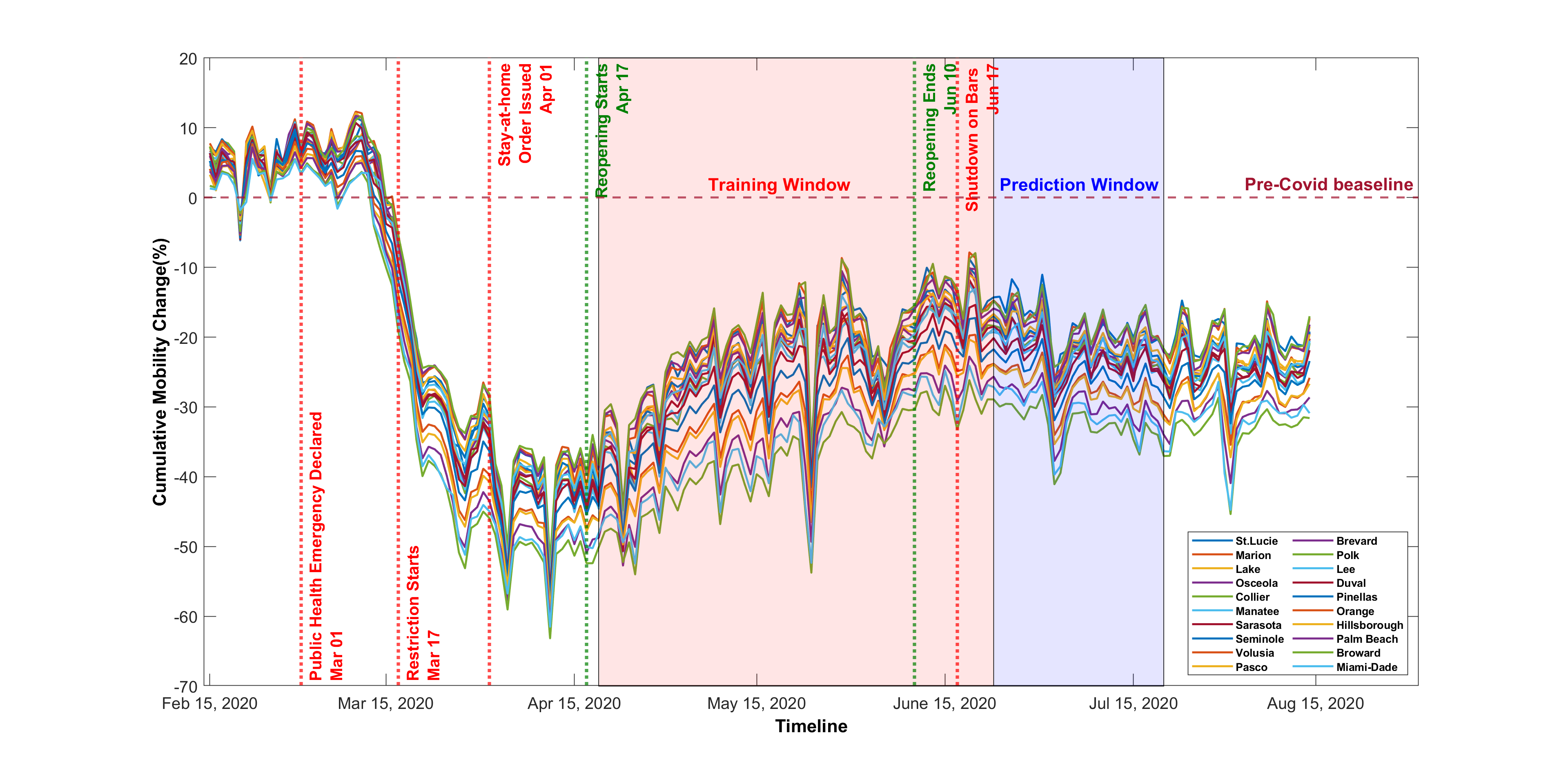}}
\caption{Percentile change of mobility in Florida from 15 February 2020 to 15 August 2020. Cumulative mobility change was calculated from \textit{Google COVID-19 Community Mobility Report} by taking the average of all modes of mobility mentioned in the data.}
\label{fig:mobility}
\end{figure*}

\subsection{Data Description} \label{data_des}

Mobility data was collected from \textit{Google COVID-19 Community Mobility Report} \cite{GoogleLLC}. This publicly available dataset contains the daily change of six different types of community mobility trends: (i) Retail and recreation ($m^{(1)}$), (ii) Grocery  and pharmacy ($m^{(2)}$), (iii) Transit stations ($m^{(3)}$), (iv) Parks ($m^{(4)}$), (v) Work stations ($m^{(5)}$), (vi) Residential ($m^{(6)}$). Let $m^{(j)}_{k,i}$ be the change in mobility on the $i^{th}$ time-frame, in the $k$-th county, for the $j$-th mode of mobility. For each time-frame $i$ and the $k$-th county, set
\begin{linenomath}
  \begin{ceqn}
   \begin{equation}
      \label{eqn:def:p_ki}
      p_{k,i} := \frac{1}{n_j} \sum_{j=1}^{n_j}m_{k,i}^{(j)}, \quad 1\leq k\leq n, \; 1\leq i\leq m ,
  \end{equation}
  \end{ceqn}
  \end{linenomath}
where $n_j$ is the number of modes of mobility. In our study we took $n_j =5$. Residential mobility was not considered while calculating mobility control input due to the underlying assumption that residential mobility is not responsible for disease spread. Besides, unlike the first five modes, residential mobility exhibited increasing trends during the pandemic.

This dataset records daily mobility data of each county regarding percentile changes concerning a normalized pre-covid baseline to adjust weekend factors. The baseline was a daily mobility average from the five weeks starting from January 3, 2020, to February 6, 2020. 
\begin{figure*}[!htbp]
\centerline{\includegraphics[width=1.2\textwidth]{\Path 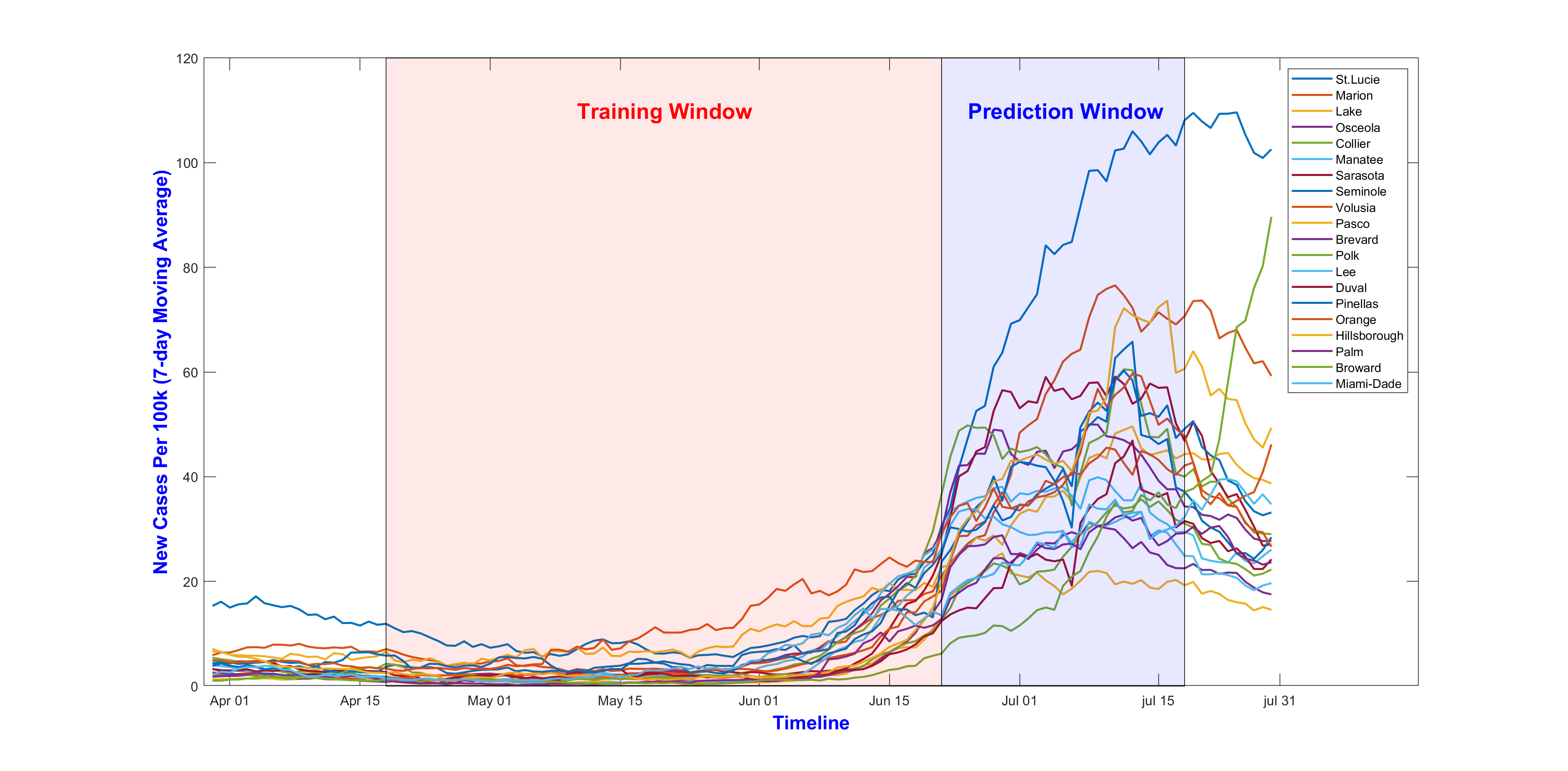}}
\caption{Moving average of daily COVID-19 case counts per 100k population in twenty most populated counties of Florida from April 01, 2020 to July 15, 2020}
\label{fig:MA}
\end{figure*}
This dataset records daily mobility data of each county regarding percentile changes concerning a normalized pre-covid baseline to adjust weekend factors. The baseline was a daily mobility average from the five weeks of January 3, 2020, to February 6, 2020. A seven-day moving average of percentile change illustrates the general mobility trend shown in Figure \ref{fig:mobility}. It also summarizes all governmental decisions. All forms of mobility exhibited strong increasing trends after the reopening decisions came in mid-April and continued until the end of June. However, partial shutdown orders on various facilities reemerged as the second wave approached in the summer. The figure shows that in Florida, during the second wave of the pandemic highest number of daily cases were reported around the third week of July 2020. However, from the fourth week of July, the trend went downwards.
\begin{figure}[!htbp]
\centerline{\includegraphics[width=0.45\textwidth]{\Path 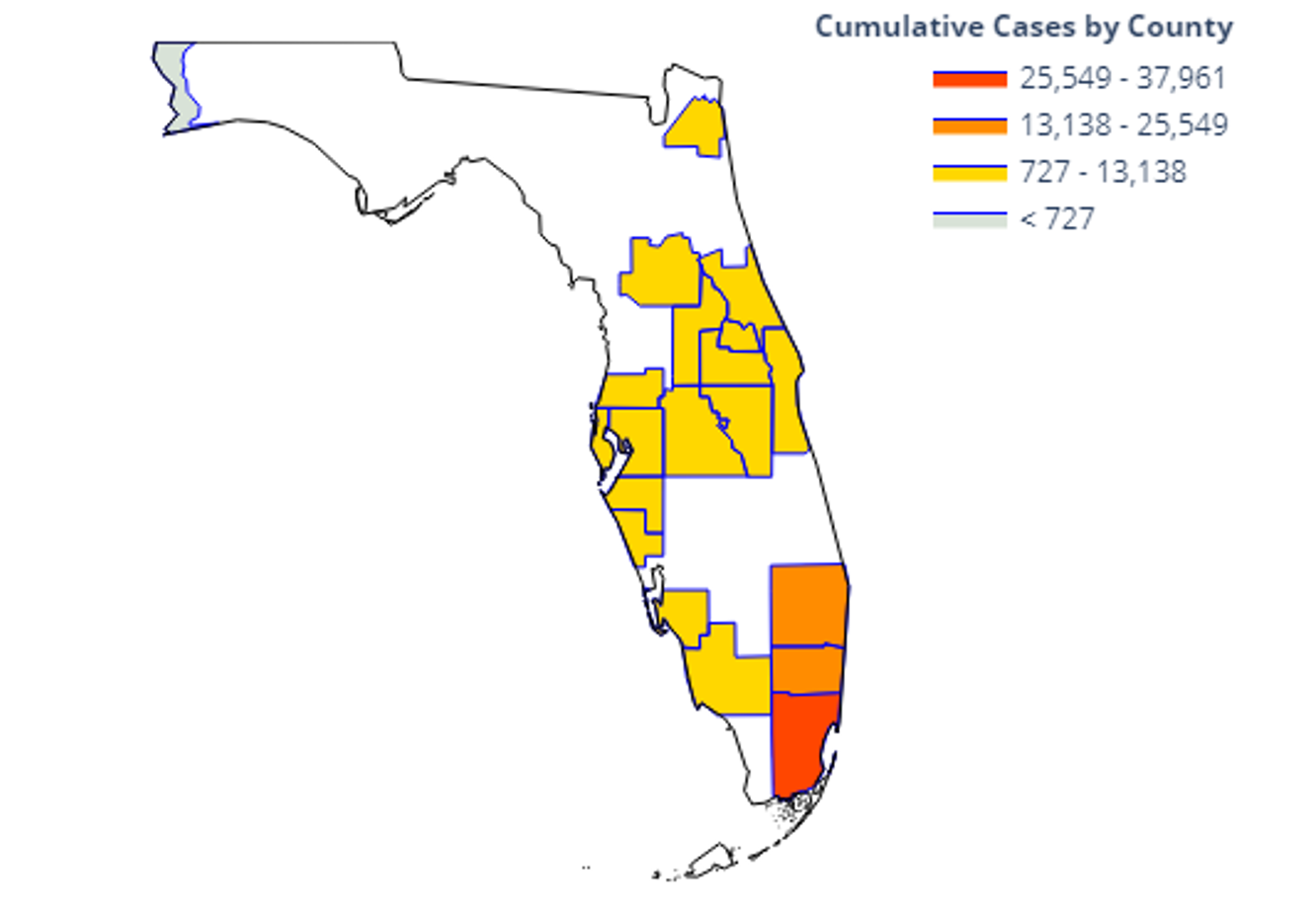}}
\caption{Spatial distribution of infection spread. Cumulative cases until 30 June 2020 are shown for the twenty most populated counties of Florida.}
\label{fig:fl}
\end{figure}
    

On the other hand, the data of county-wise COVID-19 confirmed cases in Florida was collected from USAFact public website \cite{usafacts.org_2021}. It presents daily counts of cumulative COVID cases at the county level, which is shown in the color map presented in Figure-\ref{fig:fl}.  


In this research, we considered the twenty most most populated counties in terms of total infection in Florida.  $86.3\%$ of all reported COVID-19 cases in Florida up to $01$ July $2020$ came from these twenty counties. The actual daily case counts are discrete phenomena, and thus it shows the random fluctuations in the raw data, which primarily arises due to the collective impact of weekends and reporting system of the testing process.  Rolling-average is a widely preferred solution to encounter the problem. Figure-\ref{fig:MA} shows seven day rolling average of the original data. Both Figure-\ref{fig:MA} and Figure-\ref{fig:mobility} illustrate the training and prediction window of the proposed model for visualization. The first wave of COVID-19 in Florida first wave of COVID-19 struck by the end of June. The objective of choosing the window is to investigate how accurately the model can capture that.

\subsection{Generating Control Input From Mobility Data} \label{Control Inputs}
In order to incorporate the impact of mobility into the disease spread model, we have developed a novel approach, where control input sequences are created from percentile change of mobility data obtained from google community mobility reports. For each time-frame $i$ set
\[ p_{max, i} := \max_{1\leq i \leq n} p_{k,i} , \quad p_{min, i} := \min_{1\leq \leq n} p_{k,i} , \]
and then set our control input to the $k$-th county, at the $i$-th time frame to be
\begin{equation} \label{eqn:def:u_ki}
u_{k,i} := w \frac{ p_{k,i}-p_{min, i} }{ p_{max, i} - p_{min, i} }, \quad 1\leq k\leq n, \; 1\leq i\leq m .
\end{equation}
For each time $i$, $u_i := \left( u_{k,i} \right)_{k=1}^{n} $ is the control vector that we create. The term $w$ above is a scaling factor. If $w=1$, then the components of $u_i$ lie in $[0,1]$, which could be a mismatch with the magnitude of the signals $x_k^{(h)}$. We choose $w$ by trial and error for the best forecasting results.

\begin{figure}[!htbp]
\centerline{\includegraphics[width=0.5\textwidth]{\Path 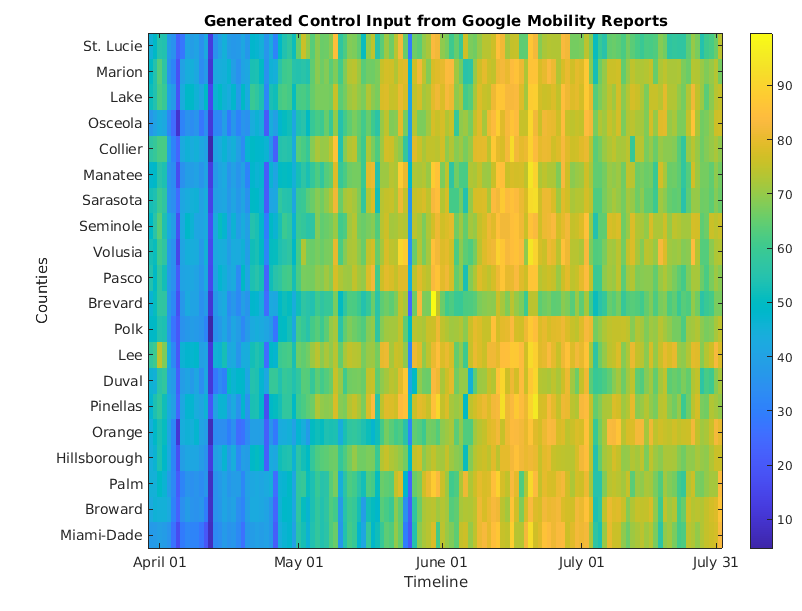}}
\caption{Control inputs are generated from \textit{Google Covid-19 Community Mobility Report}. We chose scaling factor $w = 100$ in this example}
\label{fig:5}
\end{figure} 


\section{Results and Discussion}\label{results}
In this section, we will discuss the obtained results obtained. This section is divided into two subsections. Section \ref{Model}  describes the physical significance, characteristics of the proposed linear model. Besides, we will also discuss different parameters of the linear model, i.e. effect of delay embeddings and corroboration of lead-lag relationship. In Section \ref{Forecast} we will focus on the prediction results obtained from the model.

\subsection{Characteristics of the Linear Model}\label{Model}
The proposed HDMDc-based linear model of disease spread describes the evolution of reported cumulative case counts and conjoins mobility inputs with it. The states of the dynamics consist of cumulative case counts of COVID-19 of twenty counties of Florida. The model was trained with historical cumulative case count and mobility input, while the training window spanned from 19 April 2020 to 24 June 2020. In this study, various delay coordinates were used to formulate Hankel matrix $G$ from the original data matrix $X$ and control input sequence in order to raise the dimension of observable space. It means that the observable space of the dynamics consists of previous two weeks' data. In HDMDc, the dimension of identified matrices of a nonlinear system depends on two parameters-
\begin{enumerate}
    \item Number of delay coordinates $h$
    \item Output dimension $n$ of the measurement map $\mathcal{X}$.
\end{enumerate}

\begin{figure}
\centerline{\includegraphics[width=0.6\textwidth]{\Path 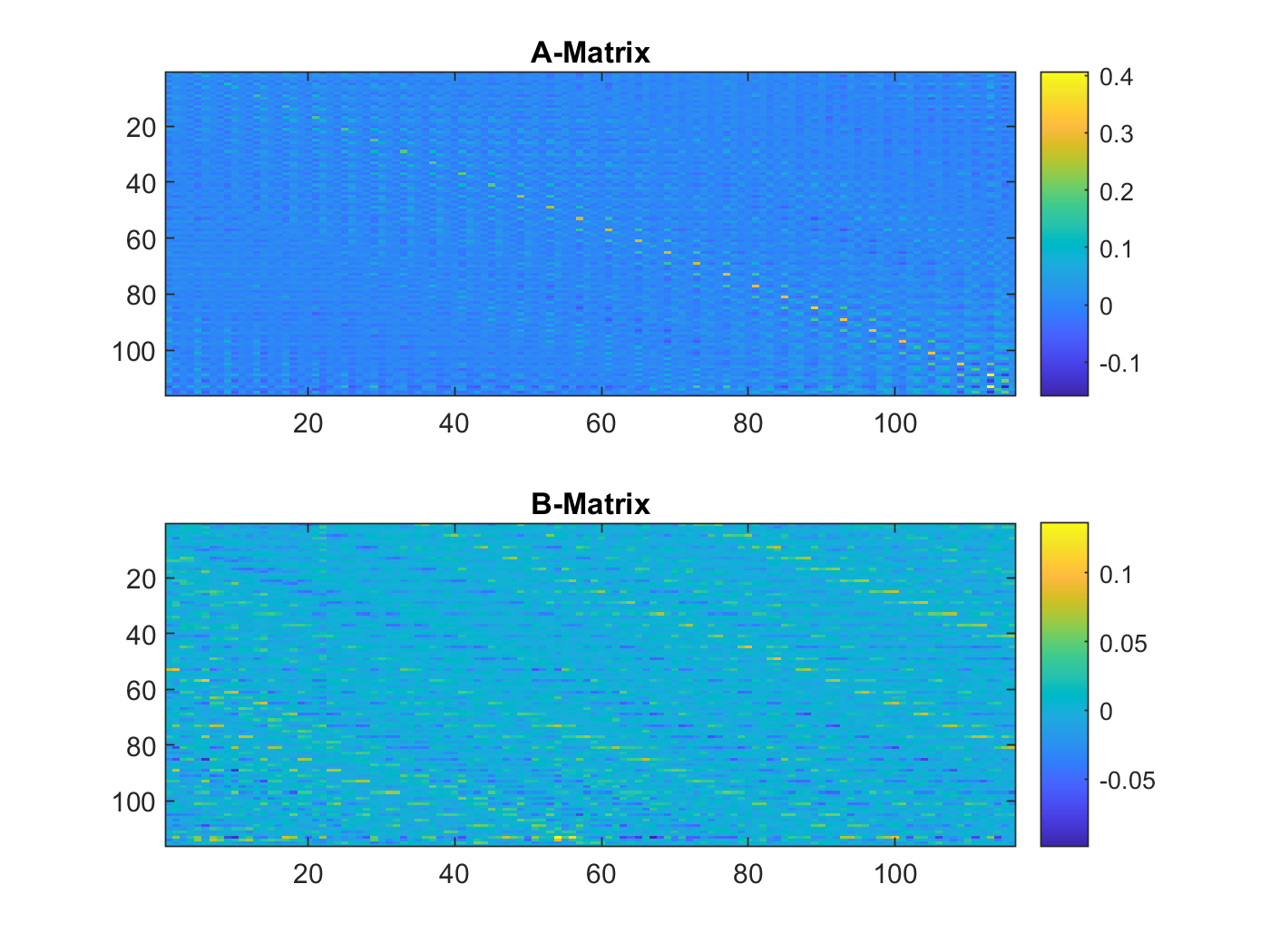}}
\caption{A and B matrices}
\label{fig:A,B}
\end{figure}

We have used 56 delay embeddings for the system identification. As a result $A,B \in \mathbb{R}^{1120\times 1120}$. As an example, Figure~\ref{fig:A,B} shows $A$ and $B$ matrices of a system consisting of four counties of Florida - Marion, Lake, Osceola, Collier. We do not show the obtained system matrices for all the twenty counties due to their large sizes which inhibits any visual interpretation.  

\begin{figure}[!htbp]
\centerline{\includegraphics[width=0.5\textwidth]{\Path 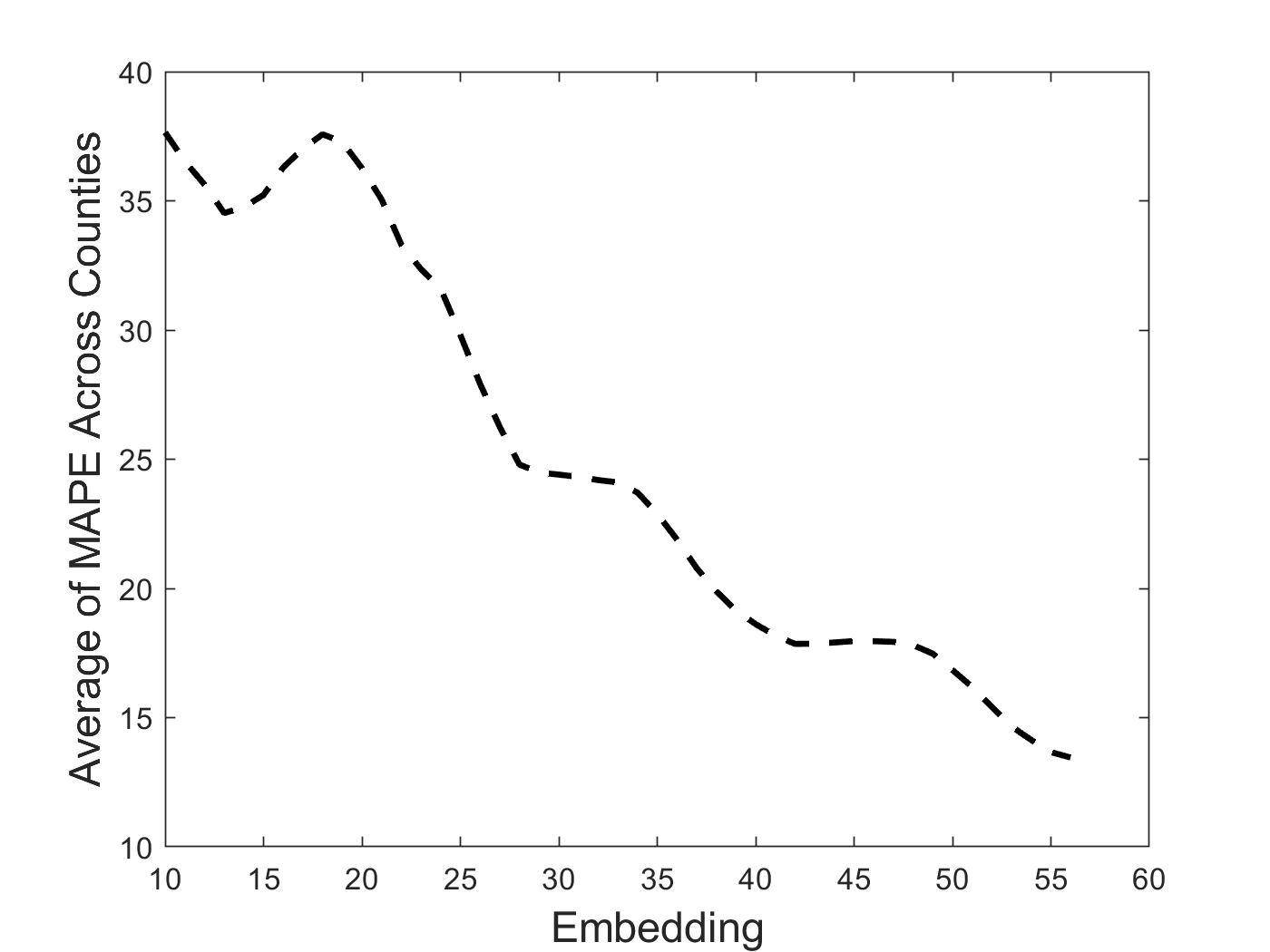}}
\caption{Average of MAPE of case count forecast across 20 counties for a four-week long prediction window vs. embedding $h$. In this plot we took moving average across four consecutive $h$ 
to understand the trend.}
\label{fig:Embedding}
\end{figure}

\begin{figure}[!htbp]
\centerline{\includegraphics[width=0.5\textwidth]{\Path 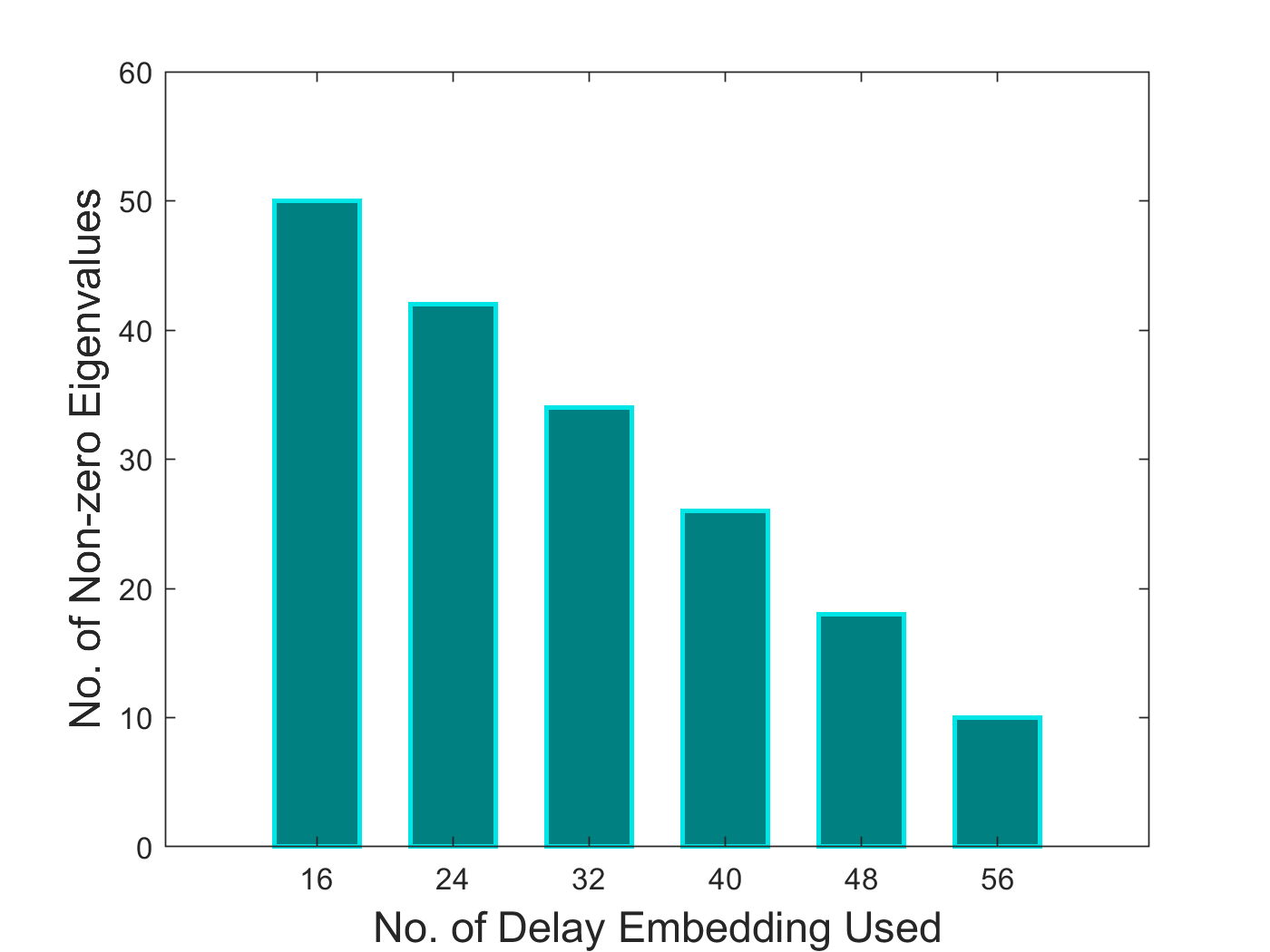}}
\caption{The relationship between number of nonzero eigenvalues of $A$ matrix with the increase of delay embedding}
\label{fig:Embedding_eig}
\end{figure}

\noindent \textbf{Choice of the number of delay embedding:} The choice of the number of delay embedding coordinates in system identification is a complex question as it depends on the nature of a system - and it is still an open research question. The choice of delay embedding in linear system identification of a nonlinear system is elaborately discussed in \cite{pan2020structure}. The relationship between delay embedding and error observed in this research is shown in Figure-\ref{fig:Embedding}. In this figure, the average prediction MAPE across all twenty counties for delay embedding ranging between $h=10$ to $h=56$ is shown. The figure shows that with the increase of delay embedding, the accuracy of prediction increases. The predictions were performed for a four-week long window.

\begin{figure*}[!htbp]
\begin{subfigure}{\textwidth}
  \centering
  \includegraphics[width=0.86\linewidth]{\Path 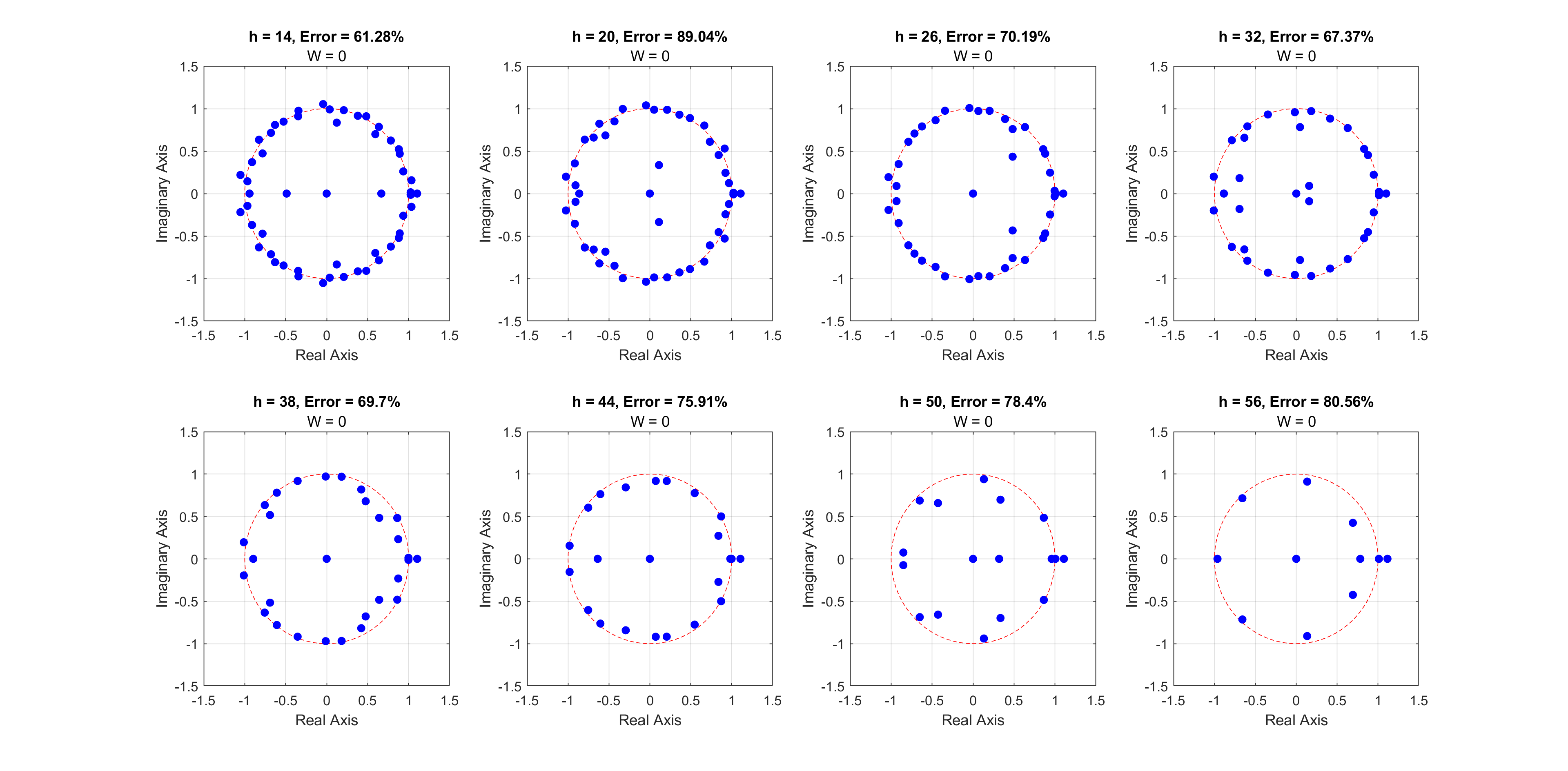}  \vspace{-7mm}
  \subcaption{}
  \label{fig:Eiga}
\end{subfigure}
\begin{subfigure}{\textwidth}
  \centering
  \includegraphics[width=0.86\linewidth]{\Path 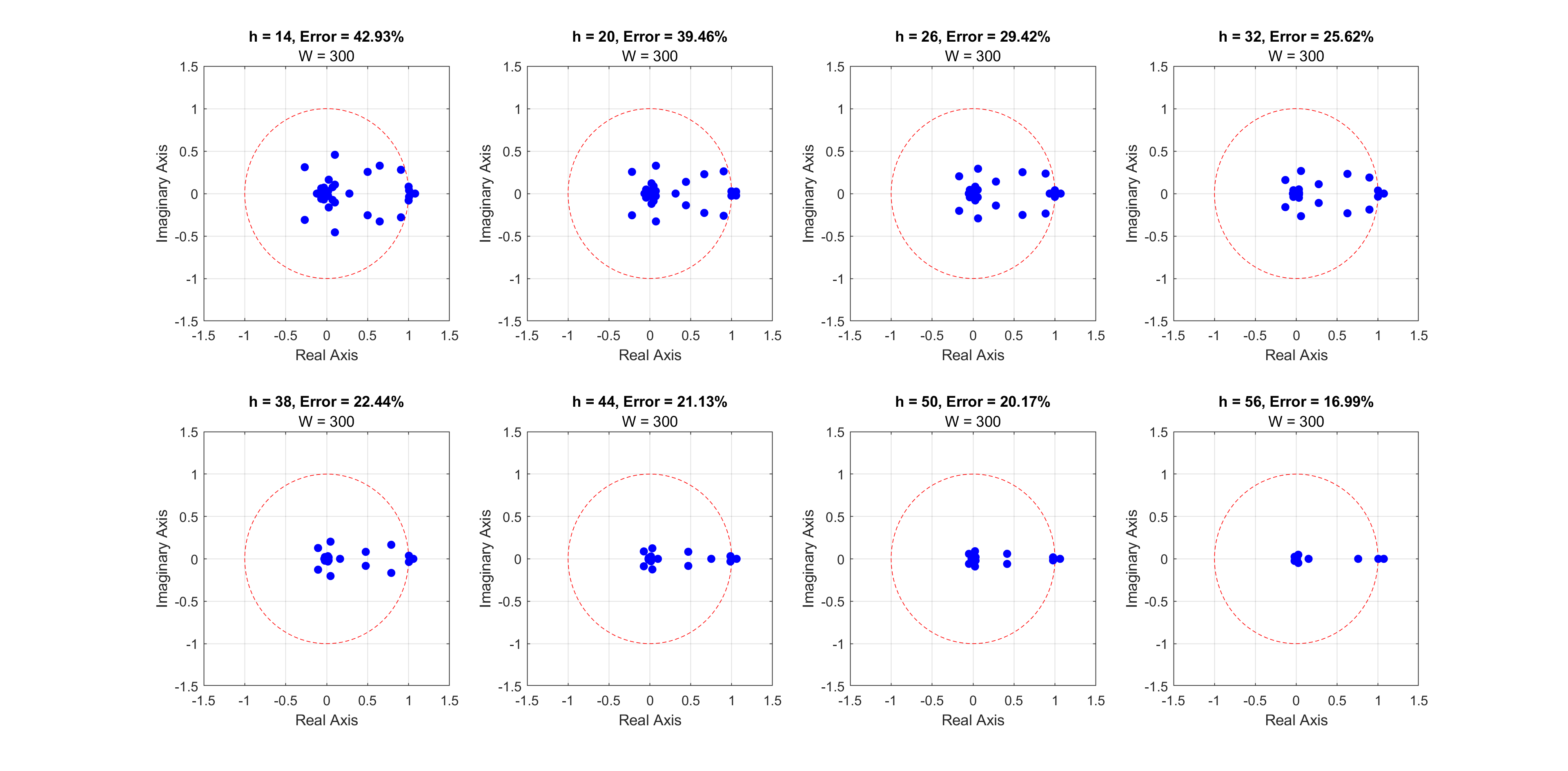}  \vspace{-7mm}
  \subcaption{}
  \label{fig:Eigb}
\end{subfigure}
 \begin{subfigure}{\textwidth}
  \centering
  \includegraphics[width=0.86\linewidth]{\Path 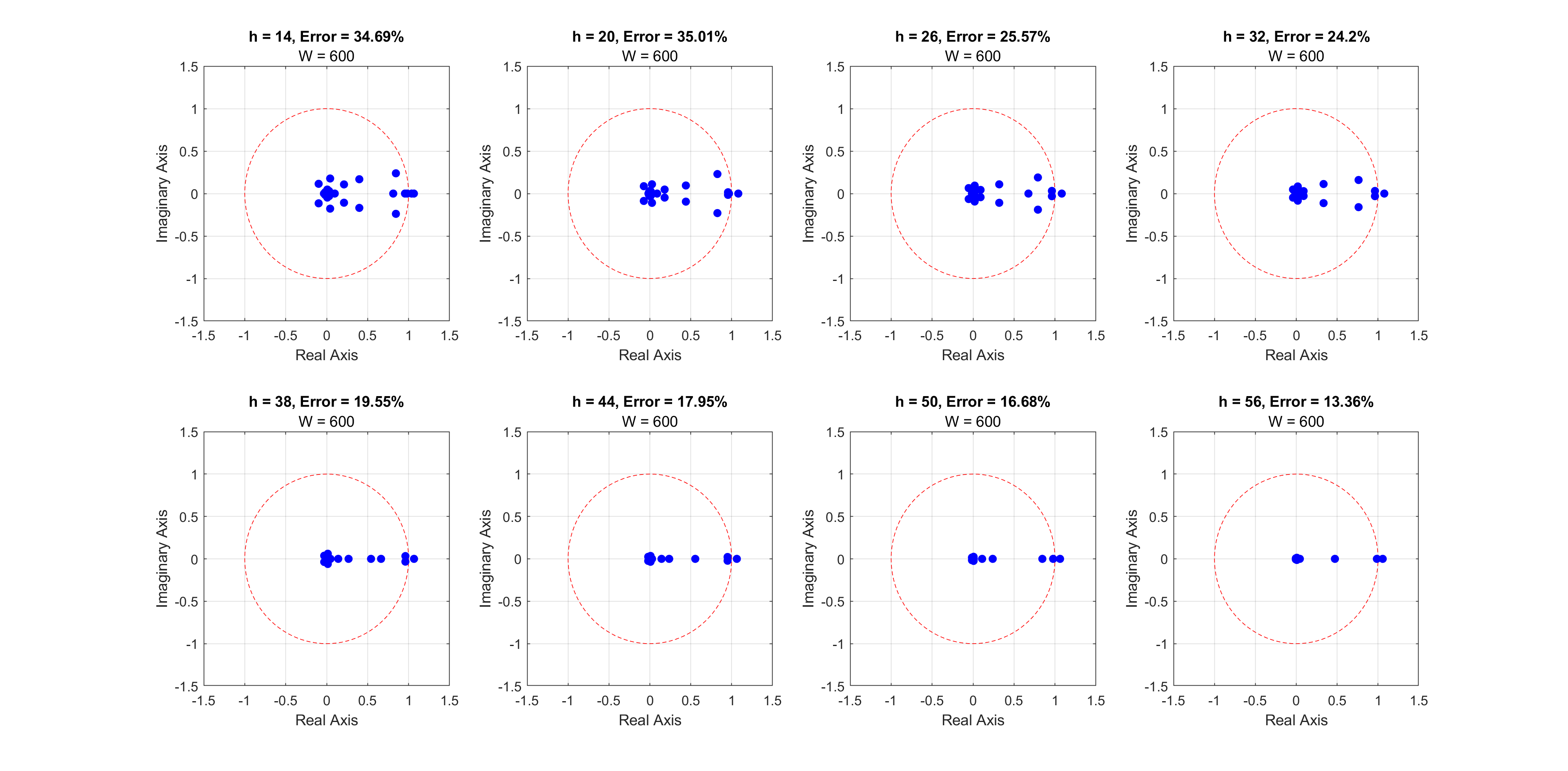}  \vspace{-7mm}
  \subcaption{}
  \label{fig:Eigc}
\end{subfigure}
\caption{Evolution of eigenvalue spectrum of system matrix $A$ for different weighting factors. Average prediction error was calculated for a 28 day prediction window across twenty counties}
\label{fig:Eig}
\end{figure*}

Figure \ref{fig:Eig} presents the impact of delay embedding coordinates in system identification accuracy. All three case figures in Figure \ref{fig:Eig} show that with the increase of the number of delay coordinate $h$, the imaginary parts of the system eigenvalues get closer to zero. Also, we observed that with the increase of $h$, the number of nonzero eigenvalues decreases. Figure \ref{fig:Embedding_eig} shows the exact relationship between $h$ and the number of nonzero eigenvalues.

Fig-\ref{fig:Eiga} shows how eigenvalues of $A$ matrix delay embedding change with the increase of delay embedding when there is no control input in the system. Similarly, fig-\ref{fig:Eigb} and, fig-\ref{fig:Eigc} shows the evolution of eigenvalue with respect to delay embedding, when weighting factor on mobility input $w = 300$ and $600$ respectively. It shows that with the increase of the weighting factor, the amplitude of eigenvalues decreases. Also note that with the increase of delay embedding and weighting factor the prediction accuracy increases.

\subsection{Prediction Results} \label{Forecast}

\begin{figure}[!h]
\centerline{\includegraphics[width=0.5\textwidth]{\Path 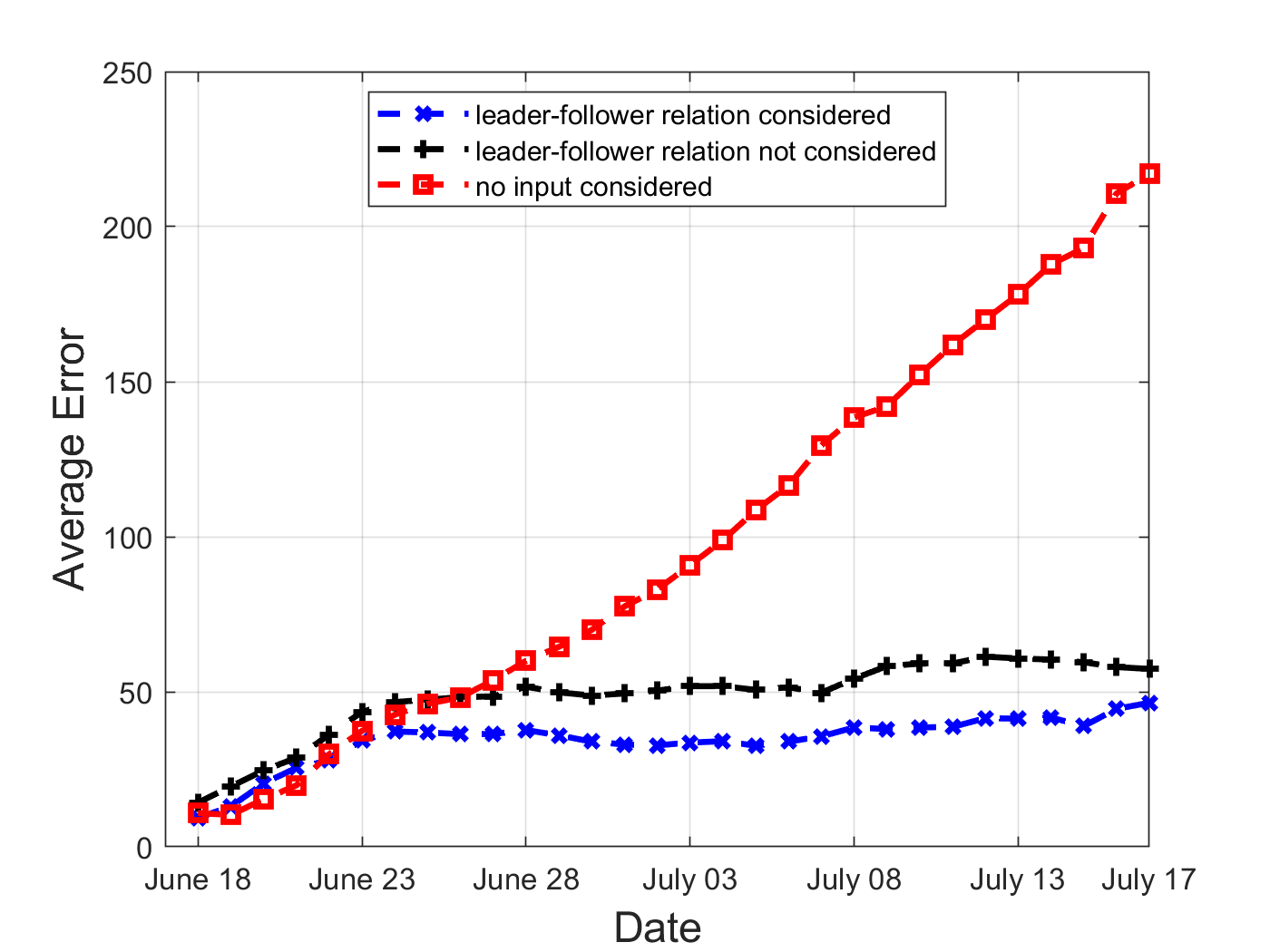}}
\caption{Case count prediction errors in three different scenarios are presented in the figure as a case study. For each day of the prediction window, we calculated the average of prediction errors across all twenty counties selected in the study.}
\label{fig:comparison}
\end{figure}

With model introduced in Section-\ref{Model} prediction on case count was made. The prediction window is stretched from June 25, 2020 to July 17,2020. We computed Mean Absolute Percentage Error (MAPE) to quantify the accuracy of the forecast. Let ${x}_{k,i}$ be the original cumulative COVID cases upto the $k^{th}$ county on any $i^{th}$ day, and $\hat{x}^{k,i}$ be the forecast. MAPE $E_{i,k}$ is defined as follows: 
\begin{equation}
 E_{k,i}=\mid\frac{{x}_{k,i} - \hat{x}_{k,i}}{{x}_{k,i}}\mid\times 100 \%     
\end{equation}
We generated control input sequences by using the strategy proposed in the section \ref{Control Inputs}. Generated control inputs are shown in Figure-\ref{fig:5}.

Mobility and the spread of infection obey a leader-follower relationship. Multiple studies such as-\cite{linka2020global,kuhl2020data,gao2021early,iacus2020human} reported that mobility trends maintain a leader-follower relationship of approximately two to three weeks with COVID-19 infection trends. The proposed model can also capture that relationship. To demonstrate that, we have reorganized the data matrix $X$ with daily cases instead of cumulative cases. Furthermore, to reflect that lead-lag relationship into the model, control inputs are shifted back 14 days, which means the current states are controlled by mobility inputs generated two weeks before.

In Figure-\ref{fig:comparison} prediction errors in  three different scenarios are plotted such as- 
\begin{itemize}
    \item case-1: considering the leader-follower relationship
    \item case-2: ignoring the leader-follower relationship
    \item case-3: ignoring all the mobility inputs
\end{itemize}

Figure~\ref{fig:comparison} shows that after leader-follower adjustment in the model, prediction results were more accurate. The figure also exhibits that the prediction results were worst when mobility was not considered as the control input. This figure supports the validity of the proposed model. First of all, it shows that mobility as an exogenous input has improved the HDMD algorithm's performance, which supports the idea of including mobility as an input into the model. Furthermore, the figure testifies that the proposed model can reflect the lag-lead relationship between COVID-19 and mobility dynamics.

\begin{table*}[!htbp]
\caption{Performance Analysis}
\centering
\begin{tabular} {ccccccc}
\hline
\multicolumn{1}{c}{} & \multicolumn{1}{c}{} & \multicolumn{3}{c}{\textbf{Prediction Error of Case Count and Effective Reproduction Number $R(t)$}}\\
\cline{1-7}
\multicolumn{1}{c}{} & \multicolumn{1}{c}{} & \multicolumn{4}{c}{\textbf{ Cumulative Case Prediction Errors}} & \multicolumn{1}{c}{\textbf{$R(t)$ Prediction Errors}}\\
\hline
\textbf{County} & \textbf{Population}   & \textbf{25 Jun'20 - 08 July'20} & \textbf{25 Jun - 14 July} & \textbf{25 June - 21 July} & \textbf{25 June - 21 July} & \textbf{01 July - 15 July}\\ 
  &   & \textbf{2020} & \textbf{2020}& \textbf{2020} & \textbf{2020} & \textbf{2020}\\
  &   & \textbf{h = 29, w = 200} & \textbf{h = 42, w = 400}& \textbf{h = 42, w = 600} & \textbf{h = 55, w = 600} & {}\\  
\hline
Miami-Dade & 2,699,428 & 4.09 & 7.49 & 5.33 & 19.35 & 18.41\\
Broward & 1,926,205 & 1.08 & 3.40 & 0.86 & 16.08 & 7.50 \\
Palm & 1,465,027 & 3.71 & 1.31 & 11.94 & 5.55 & 11.43 \\
Hillsborough & 1,422,278 & 3.39 & 12.47 & 25.83 & 3.00 & 14.15 \\
Orange & 1,349,746 & 5.57 & 21.52 & 31.63 & 3.62 & 15.10 \\
Pinellas & 964,666 & 12.37 & 24.51 & 47.67 & 17.49 & 4.44 \\
Duval & 936,186 & 8.61 & 1.05 & 1.22 & 21.81 & 6.92 \\ 
St.Lucie & 312,947 & 5.47 & 8.13 & 1.22 & 13.09 & 11.58  \\
Marion & 353,526 & 2.37 & 2.60 & 9.14 & 10.46 & 4.92  \\
Lake & 345,867 & 3.20 & 9.17 & 19.50 & 8.82 & 9.12 \\
Osceola & 351,955 & 2.10 & 11.92 & 15.13 & 11.22 & 5.96  \\
Collier & 371,453 & 1.65 & 4.19  & 7.21 & 13.97 & 10.74\\
Manatee & 384,213 & 15.13 & 35.45 & 57.57 & 19.77 & 4.90 \\
Sarasota & 419,496 & 0.57  & 6.49  & 3.92 & 17.31 & 15.75 \\
Seminole & 461,402 & 10.41 & 19.20 & 27.76 & 3.82 & 11.37 \\
Volusia & 536,487 & 2.56  & 0.18  & 6.98 & 8.76 & 9.00\\
Pasco & 524,602 & 9.79  & 6.24  & 11.83 & 30.70 & 6.41 \\
Brevard & 585,507 & 1.52 & 4.11  & 24.31 & 6.31 & 15.61 \\
Polk & 686,218 & 1.04  & 4.19  & 16.53 & 32.41 & 16.66 \\
Lee & 737,468 & 0.11  & 6.77  & 8.83 & 2.32  & 12.68 \\\hline
Average & & \textbf{4.74 \%} & \textbf{9.52 \%} & \textbf{16.72 \%} & \textbf{13.29\%} & \textbf{10.63 \%} \\ \hline
\end{tabular}
\label{table:1}
\end{table*}

\begin{figure*}[!htbp]
\centerline{\includegraphics[width=1.2\textwidth]{\Path 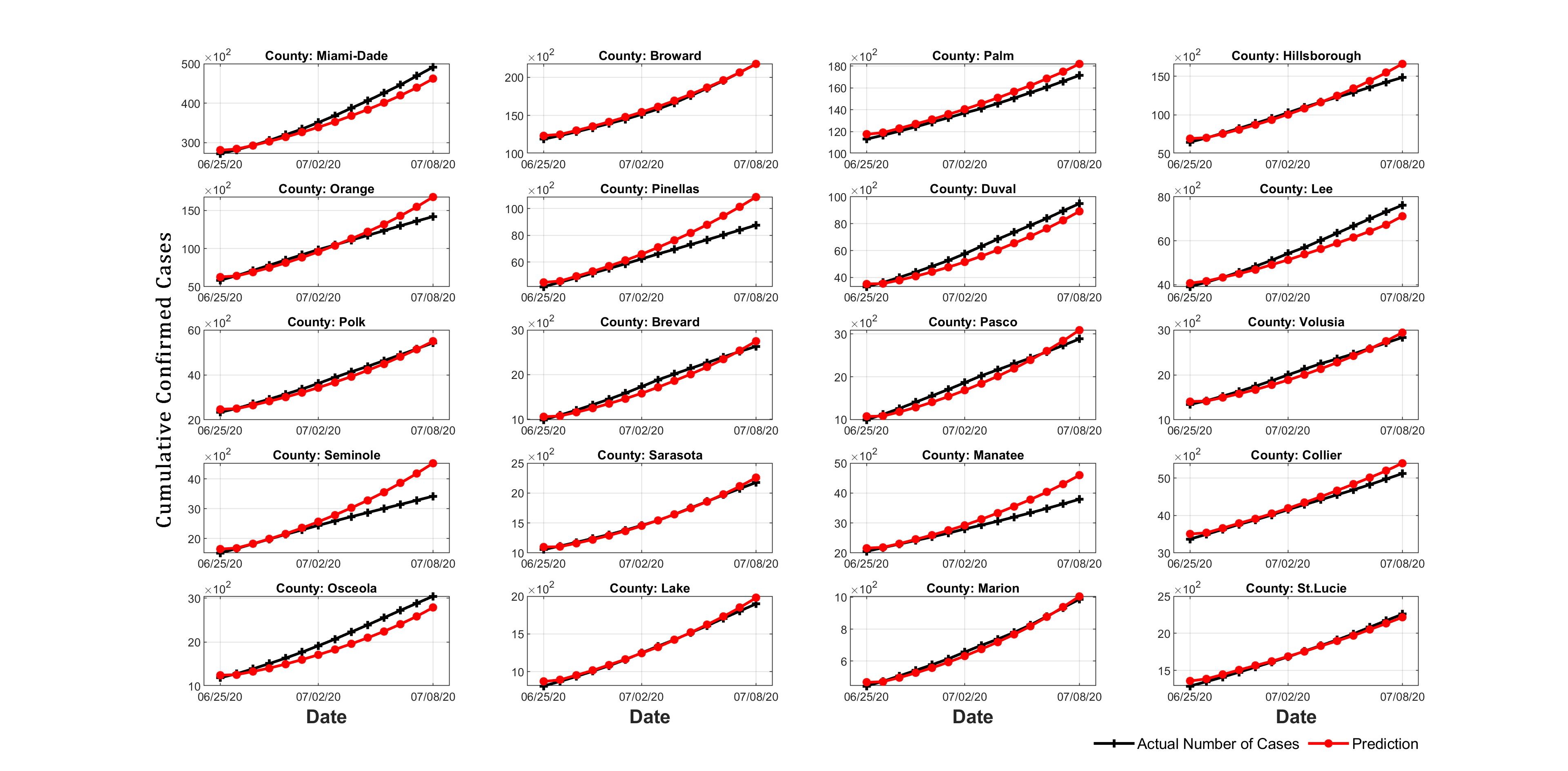}}
\caption{Cumulative case prediction for two weeks window - 25 June to 02 July, 2020. We selected $h=29$ and $w=200$ for the prediction}
\label{fig:6}
\end{figure*}

\begin{figure*}[!htbp]
\centerline{\includegraphics[width=1.2\textwidth]{\Path 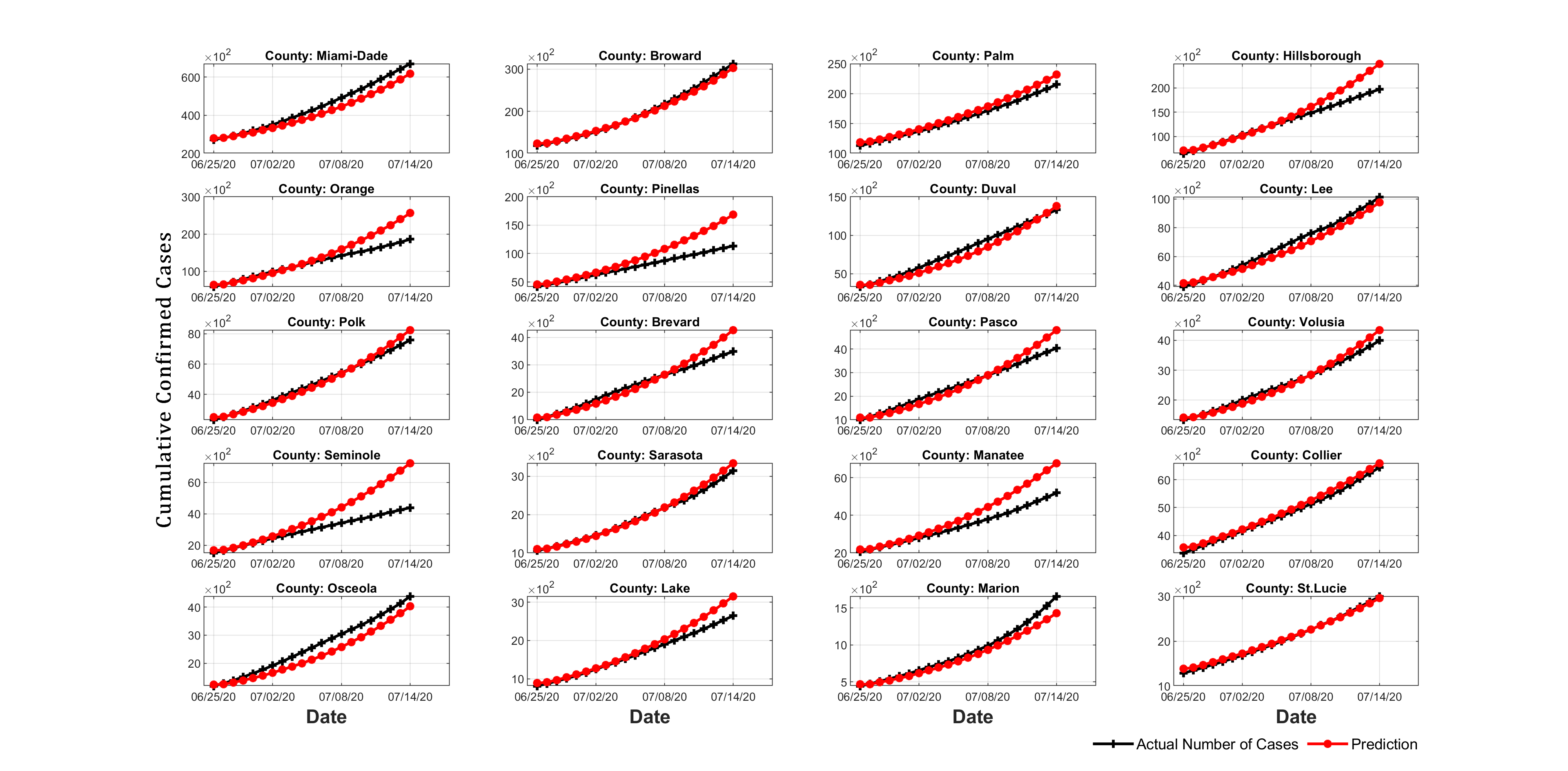}}
\caption{Cumulative case prediction for three weeks window - 25 June to 08 July, 2020. We selected $h=42$ and $w=400$ for the prediction}
\label{fig:14}
\end{figure*}

\begin{figure*}[!htbp]
\centerline{\includegraphics[width=1.2\textwidth]{\Path 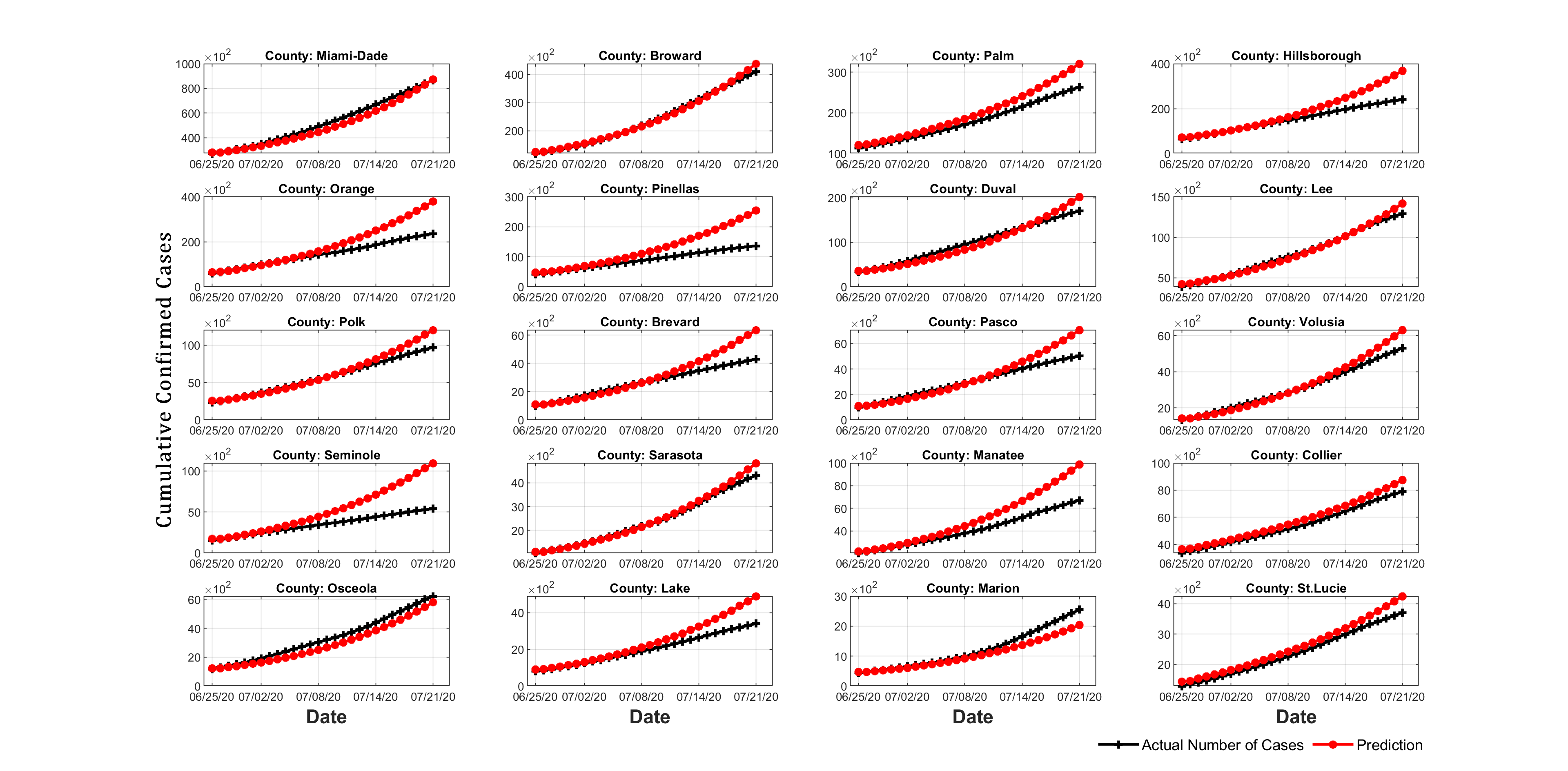}}
\caption{Cumulative case prediction for four weeks window - 25 June to 15 July, 2020. We selected $h=42$ and $w=600$ for the prediction}
\label{fig:15}
\end{figure*}

Although prediction is not a primary application of DMD-type algorithms, we made forecasts to demonstrate the model's validity. Figure \ref{fig:6} and Figure \ref{fig:14}, Figure \ref{fig:15}, show prediction results respectively for two-week and three-week and four-week window for all the twenty counties we selected for this study. Table \ref{table:1} illustrates average prediction errors. We calculated MAPE across the prediction window, and we define it as the prediction error for a specific county for a specific prediction window. We also included the number of delay embedding $h$ and weight $w$ on mobility used in each prediction window. 

In the two-week window, overall MAPE prediction error across twenty counties was less than 5\%. Prediction performance in Lee county was found to be best as the model could Predict total cases with an estimated error of only 0.1\%. On the other hand, the model performed worst for Manatee county, where the prediction error was $15.13\%.$  

In the three-week window, overall MAPE prediction error across twenty counties was less than 10\%.
In this window, the model's prediction performance for Volusia county was the best as it could predict total cases with an estimated error of only 0.18\%, while the performance was the worst for Manatee county. For the latter error was 35.45$\%$.

The table also shows that overall prediction performance was best for the two-week prediction window, which means that after weeks, HDMDc based prediction performance exacerbates. One possible explanation of deterioration happens because HDMDc is a locally linear approximation of a complex dynamical system. Intuitively it makes sense that for more extended periods, this locally linear approximation will be less valid.

In the four-week window, we considered two different delay embedding $h = 42$ and $56$. For $h = 42$ overall prediction error of the model was worse than that of $h = 56$. However, we observed some interesting patterns here. Setting $h = 42$, it was found that for five counties, prediction errors were more than $25\%$. For Manatee county, the model performed the worst with an error of 57.57$\%$. On the other hand, when we chose $h = 56$, only two counties exhibited prediction error more than $25\%$, and the worst prediction result was found in Polk county with an error of $32.41\%$. However, prediction performance drastically deteriorated for the two most populated and COVID-infested counties of Florida, namely- Miami-Dade and Broward. It implies that although increasing embedding improves the model's overall performance, it can deteriorate prediction accuracy for some counties. Hence, careful selection of delay embedding is necessary, and it should be selected as per the model's objective. Although we have chosen some specific number of delay embedding in this study i.e. $h = 29, 42$, and $56$, there would not have been much difference, if we chose $h = 31$ instead of $h = 29$, or $h = 52$ instead of $h = 56$. The underlying rationale can be understood from Figure \ref{fig:Embedding}. The figure shows that for a particular band of $h$, the prediction error remains almost the same. However, errors tend to decrease with the increase of $h$.     

We studied the role of delay embedding $h$ and scaling factor $w$ in forecasting performance. We observed that increasing the number of embedding and scaling factor improves forecasting performances for longer prediction windows. When the number of embedding increases, the model can train itself with a richer set of historical values, which might improve its accuracy for a longer prediction window; whereas for a shorter window, the model can adjust itself with a lesser number of delay embedding. The results also show that forecasting performance improves when we assign heavier scaling factors on control inputs for longer prediction windows.       

\begin{figure*}[!htbp]
\centerline{\includegraphics[width=1.2\textwidth]{\Path 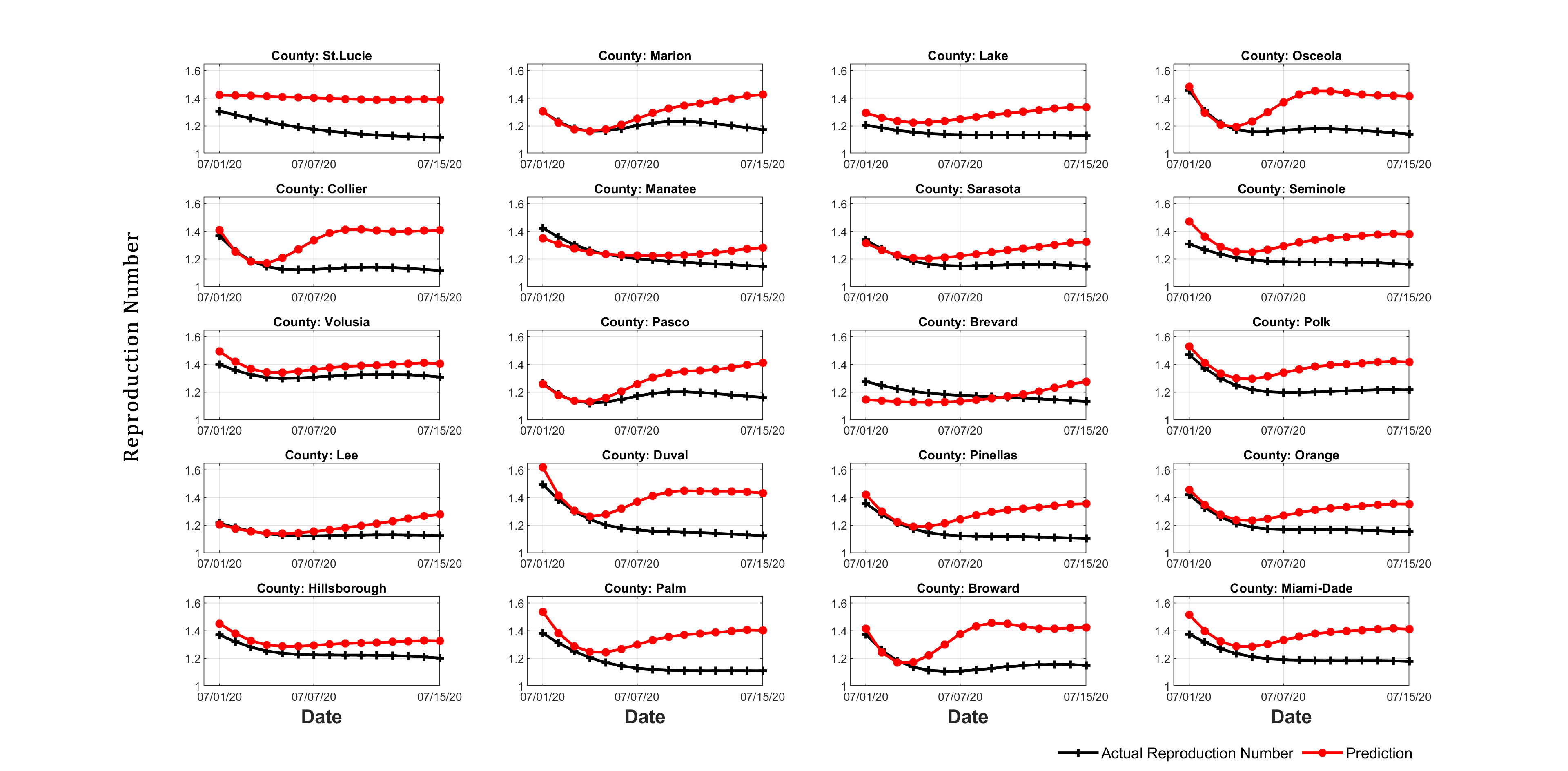}}
\caption{Reproduction number prediction - 01 July to July 15, 2020}
\label{fig:rep}
\end{figure*}   



In order to validate the proposed model, further reproduction number of COVID-19 was estimated from the infection case counts forecast. By definition, reproduction number is the average fractional increase of the number of new cases with respect to the existing cases during the infectious period. For example, if the infectious period of COVID-19 is considered 34 days and the total number of cases are increased 2.5 times during those 34 days, the reproduction number is $R_{0}$ is 2.5. However, disease spread is a dynamic incident; hence $R_{0}$ keeps evolving with time. We estimated effective reproduction number $R_{t}$ by using the predicted case counts obtained from the model. For the computation we used a python module called epyestim \cite{hilfiker_josi}. The county-wise prediction errors are shown in the last column of the Table \ref{table:1}, while county-wise prediction results are shown in \ref{fig:rep}. We observe that in Figure \ref{fig:rep} in most of the cases, estimated results diverged from the original $R$ after the first week of the window. However, the overall prediction error was $10.63\%$, and the difference between actual and predicted $R(t)$ was around $0.2$ even in the worst cases. It implies that the proposed model is also useful for forecasting the dynamic reproduction number, which is an important parameter for any epidemiological modeling.




\section{CONCLUSIONS}

This paper developed a novel framework for data assimilation and uncovering interconnections between human activity and mobility (HAM) and disease spread. The study exploits recent advances in Koopman operator theory to understand the relationship between epidemic dynamics and HAM. The proposed framework results in a locally linear model (unlike SEIR and other traditional mechanistic models) for disease spread where HAM acts as an external influence. It is a purely data-driven model which does not need any parameter estimation. Although the approach is data-driven, the framework should not be confused with black-box machine learning models. A case study was performed for COVID-19 spread in Florida. The obtained linear model successfully predicted new cases and $R_0$ over the next few weeks window. There are a few shortcomings in the study which can be improved in future studies. Firstly, the study considers local HAM data (i.e., within each county) - and does not consider inter-county HAM data. In other words, it does not consider the impact of global mobility. The HAM is a complex phenomenon, and it is challenging to consider every aspect of it in the modeling. For example, HAMs such as private parties, spring break traveling, usage of face masks, long-term travel, etc. Secondly, the study was focused on the early phase of COVID-19 spread. It will be interesting to study how are HAM and disease spread were related during the later stages of the pandemic.


FloatBarrier


\newcommand{\BIBdecl}{\setlength{\itemsep}{0.25 em}}
\bibliography{CCTA_Ref, Shuddho_references}

\begin{thebibliography}{10}
\providecommand{\url}[1]{#1}
\csname url@samestyle\endcsname
\providecommand{\newblock}{\relax}
\providecommand{\bibinfo}[2]{#2}
\providecommand{\BIBentrySTDinterwordspacing}{\spaceskip=0pt\relax}
\providecommand{\BIBentryALTinterwordstretchfactor}{4}
\providecommand{\BIBentryALTinterwordspacing}{\spaceskip=\fontdimen2\font plus
\BIBentryALTinterwordstretchfactor\fontdimen3\font minus
  \fontdimen4\font\relax}
\providecommand{\BIBforeignlanguage}[2]{{%
\expandafter\ifx\csname l@#1\endcsname\relax
\typeout{** WARNING: IEEEtran.bst: No hyphenation pattern has been}%
\typeout{** loaded for the language `#1'. Using the pattern for}%
\typeout{** the default language instead.}%
\else
\language=\csname l@#1\endcsname
\fi
#2}}
\providecommand{\BIBdecl}{\relax}
\BIBdecl

\bibitem{balcan2009multiscale}
D.~Balcan, V.~Colizza, B.~Gon{\c{c}}alves, H.~Hu, J.~J. Ramasco, and
  A.~Vespignani, ``Multiscale mobility networks and the spatial spreading of
  infectious diseases,'' \emph{Proceedings of the National Academy of
  Sciences}, vol. 106, no.~51, pp. 21\,484--21\,489, 2009.

\bibitem{colizza2006role}
V.~Colizza, A.~Barrat, M.~Barth{\'e}lemy, and A.~Vespignani, ``The role of the
  airline transportation network in the prediction and predictability of global
  epidemics,'' \emph{Proceedings of the National Academy of Sciences}, vol.
  103, no.~7, pp. 2015--2020, 2006.

\bibitem{quaranta2020understanding}
G.~Quaranta, G.~Formica, J.~T. Machado, W.~Lacarbonara, and S.~F. Masri,
  ``Understanding covid-19 nonlinear multi-scale dynamic spreading in italy,''
  \emph{Nonlinear Dynamics}, vol. 101, no.~3, pp. 1583--1619, 2020.

\bibitem{linka2020global}
K.~Linka, A.~Goriely, and E.~Kuhl, ``Global and local mobility as a barometer
  for covid-19 dynamics,'' \emph{medRxiv}, 2020.

\bibitem{bernoulli1760essai}
D.~Bernoulli, ``Essai d'une nouvelle analyse de la mortalit{\'e} caus{\'e}e par
  la petite v{\'e}role, et des avantages de l'inoculation pour la
  pr{\'e}venir,'' \emph{Histoire de l'Acad., Roy. Sci.(Paris) avec Mem}, pp.
  1--45, 1760.

\bibitem{hethcote2000mathematics}
H.~W. Hethcote, ``The mathematics of infectious diseases,'' \emph{SIAM review},
  vol.~42, no.~4, pp. 599--653, 2000.

\bibitem{brauer2019mathematical}
F.~Brauer, C.~Castillo-Chavez, and Z.~Feng, \emph{Mathematical models in
  epidemiology}.\hskip 1em plus 0.5em minus 0.4em\relax Springer, 2019,
  vol.~32.

\bibitem{rahimi2021review}
I.~Rahimi, F.~Chen, and A.~H. Gandomi, ``A review on covid-19 forecasting
  models,'' \emph{Neural Computing and Applications}, pp. 1--11, 2021.

\bibitem{james2021use}
L.~P. James, J.~A. Salomon, C.~O. Buckee, and N.~A. Menzies, ``The use and
  misuse of mathematical modeling for infectious disease policymaking: Lessons
  for the covid-19 pandemic,'' \emph{Medical Decision Making}, p.
  0272989X21990391, 2021.

\bibitem{holmdahl2020wrong}
I.~Holmdahl and C.~Buckee, ``Wrong but useful—what covid-19 epidemiologic
  models can and cannot tell us,'' \emph{New England Journal of Medicine}, vol.
  383, no.~4, pp. 303--305, 2020.

\bibitem{kuhl2020data}
E.~Kuhl, ``Data-driven modeling of covid-19—lessons learned,'' \emph{Extreme
  Mechanics Letters}, p. 100921, 2020.

\bibitem{peirlinck2020outbreak}
M.~Peirlinck, K.~Linka, F.~S. Costabal, and E.~Kuhl, ``Outbreak dynamics of
  covid-19 in china and the united states,'' \emph{Biomechanics and modeling in
  mechanobiology}, p.~1, 2020.

\bibitem{he2020seir}
S.~He, Y.~Peng, and K.~Sun, ``Seir modeling of the covid-19 and its dynamics,''
  \emph{Nonlinear Dynamics}, vol. 101, no.~3, pp. 1667--1680, 2020.

\bibitem{cobey2020modeling}
S.~Cobey, ``Modeling infectious disease dynamics,'' \emph{Science}, vol. 368,
  no. 6492, pp. 713--714, 2020.

\bibitem{bajardi2011human}
P.~Bajardi, C.~Poletto, J.~J. Ramasco, M.~Tizzoni, V.~Colizza, and
  A.~Vespignani, ``Human mobility networks, travel restrictions, and the global
  spread of 2009 h1n1 pandemic,'' \emph{PloS one}, vol.~6, no.~1, p. e16591,
  2011.

\bibitem{khan2009spread}
K.~Khan, J.~Arino, W.~Hu, P.~Raposo, J.~Sears, F.~Calderon, C.~Heidebrecht,
  M.~Macdonald, J.~Liauw, A.~Chan \emph{et~al.}, ``Spread of a novel influenza
  a (h1n1) virus via global airline transportation,'' \emph{New England journal
  of medicine}, vol. 361, no.~2, pp. 212--214, 2009.

\bibitem{cauchemez2011role}
S.~Cauchemez, A.~Bhattarai, T.~L. Marchbanks, R.~P. Fagan, S.~Ostroff, N.~M.
  Ferguson, D.~Swerdlow, P.~H.~W. Group \emph{et~al.}, ``Role of social
  networks in shaping disease transmission during a community outbreak of 2009
  h1n1 pandemic influenza,'' \emph{Proceedings of the National Academy of
  Sciences}, vol. 108, no.~7, pp. 2825--2830, 2011.

\bibitem{herrera2011multiple}
M.~A. Herrera-Valdez, M.~Cruz-Aponte, and C.~Castillo-Chavez, ``Multiple
  outbreaks for the same pandemic: Local transportation and social distancing
  explain the different" waves" of a-h1n1pdm cases observed in m{\'e}xico
  during 2009,'' \emph{Mathematical Biosciences \& Engineering}, vol.~8, no.~1,
  p.~21, 2011.

\bibitem{espinoza2020mobility}
B.~Espinoza, C.~Castillo-Chavez, and C.~Perrings, ``Mobility restrictions for
  the control of epidemics: When do they work?'' \emph{Plos one}, vol.~15,
  no.~7, p. e0235731, 2020.

\bibitem{race1995some}
P.~Race, ``Some further consideration of the plague in eyam, 1665/6,''
  \emph{Local population studies}, vol.~54, pp. 56--65, 1995.

\bibitem{kohn2007encyclopedia}
G.~C. Kohn, \emph{Encyclopedia of plague and pestilence: from ancient times to
  the present}.\hskip 1em plus 0.5em minus 0.4em\relax Infobase Publishing,
  2007.

\bibitem{cetron2005public}
M.~Cetron and J.~Landwirth, ``Public health and ethical considerations in
  planning for quarantine.'' \emph{The Yale journal of biology and medicine},
  vol.~78, no.~5, p. 329, 2005.

\bibitem{baroyan1971computer}
O.~Baroyan, L.~Rvachev, U.~Basilevsky, V.~Ermakov, K.~Frank, M.~Rvachev, and
  V.~Shashkov, ``Computer modelling of influenza epidemics for the whole
  country (ussr),'' \emph{Advances in Applied Probability}, vol.~3, no.~2, pp.
  224--226, 1971.

\bibitem{rvachev1985mathematical}
L.~A. Rvachev and I.~M. Longini~Jr, ``A mathematical model for the global
  spread of influenza,'' \emph{Mathematical biosciences}, vol.~75, no.~1, pp.
  3--22, 1985.

\bibitem{xiong2020mobile}
C.~Xiong, S.~Hu, M.~Yang, W.~Luo, and L.~Zhang, ``Mobile device data reveal the
  dynamics in a positive relationship between human mobility and covid-19
  infections,'' \emph{Proceedings of the National Academy of Sciences}, vol.
  117, no.~44, pp. 27\,087--27\,089, 2020.

\bibitem{wang2020using}
H.~Wang and N.~Yamamoto, ``Using a partial differential equation with google
  mobility data to predict covid-19 in arizona,'' \emph{Mathematical
  Biosciences and Engineering}, vol.~17, no.~5, 2020.

\bibitem{zengspatial}
C.~Zeng, J.~Zhang, Z.~Li, X.~Sun, B.~Olatosi, S.~Weissman, and X.~Li,
  ``Spatial-temporal relationship between population mobility and covid-19
  outbreaks in south carolina: A time series forecasting analysis,''
  \emph{medRxiv: the preprint server for health sciences}, pp. 2021--01, 2021.

\bibitem{hubig}
S.~Hu, C.~Xiong, M.~Yang, H.~Younes, W.~Luo, and L.~Zhang, ``A big-data driven
  approach to analyzing and modeling human mobility trend under
  non-pharmaceutical interventions during covid-19 pandemic,''
  \emph{Transportation Research Part C: Emerging Technologies}, p. 102955,
  2021.

\bibitem{iacus2020human}
S.~M. Iacus, C.~Santamaria, F.~Sermi, S.~Spyratos, D.~Tarchi, and M.~Vespe,
  ``Human mobility and covid-19 initial dynamics,'' \emph{Nonlinear Dynamics},
  vol. 101, no.~3, pp. 1901--1919, 2020.

\bibitem{linka2020outbreak}
K.~Linka, M.~Peirlinck, F.~Sahli~Costabal, and E.~Kuhl, ``Outbreak dynamics of
  covid-19 in europe and the effect of travel restrictions,'' \emph{Computer
  Methods in Biomechanics and Biomedical Engineering}, pp. 1--8, 2020.

\bibitem{proctor2016dynamic}
J.~L. Proctor, S.~L. Brunton, and J.~N. Kutz, ``Dynamic mode decomposition with
  control,'' \emph{SIAM Journal on Applied Dynamical Systems}, vol.~15, no.~1,
  pp. 142--161, 2016.

\bibitem{fazel2013hankel}
M.~Fazel, T.~K. Pong, D.~Sun, and P.~Tseng, ``Hankel matrix rank minimization
  with applications to system identification and realization,'' \emph{SIAM
  Journal on Matrix Analysis and Applications}, vol.~34, no.~3, pp. 946--977,
  2013.

\bibitem{avila2020data}
A.~Avila and I.~Mezi{\'c}, ``Data-driven analysis and forecasting of highway
  traffic dynamics,'' \emph{Nature communications}, vol.~11, no.~1, pp. 1--16,
  2020.

\bibitem{ling2018koopman}
E.~Ling, L.~Ratliff, and S.~Coogan, ``Koopman operator approach for instability
  detection and mitigation in signalized traffic,'' in \emph{2018 21st
  International Conference on Intelligent Transportation Systems (ITSC)}.\hskip
  1em plus 0.5em minus 0.4em\relax IEEE, 2018, pp. 1297--1302.

\bibitem{boskic2020koopman}
L.~Boskic, C.~N. Brown, and I.~Mezi{\'c}, ``Koopman mode analysis on thermal
  data for building energy assessment,'' \emph{Advances in Building Energy
  Research}, pp. 1--15, 2020.

\bibitem{sauer1991embedology}
T.~Sauer, J.~A. Yorke, and M.~Casdagli, ``Embedology,'' \emph{Journal of
  statistical Physics}, vol.~65, no.~3, pp. 579--616, 1991.

\bibitem{hunt1992prevalence}
B.~R. Hunt, T.~Sauer, and J.~A. Yorke, ``Prevalence: a translation-invariant
  “almost every” on infinite-dimensional spaces,'' \emph{Bulletin of the
  American mathematical society}, vol.~27, no.~2, pp. 217--238, 1992.

\bibitem{erichson2019compressed}
N.~B. Erichson, S.~L. Brunton, and J.~N. Kutz, ``Compressed dynamic mode
  decomposition for background modeling,'' \emph{Journal of Real-Time Image
  Processing}, vol.~16, no.~5, pp. 1479--1492, 2019.

\bibitem{DasGiannakis_delay_2019}
S.~Das and D.~Giannakis, ``Delay-coordinate maps and the spectra of {K}oopman
  operators,'' \emph{J. Stat. Phys.}, vol. 175, p. 1107–1145, 2019.

\bibitem{DGJ_compactV_2018}
D.~Giannakis, S.~Das, and J.~Slawinska, ``Reproducing kernel {H}ilbert space
  compactification of unitary evolution groups,'' \emph{Appl. Comput. Harmon.
  Anal.}, vol.~54, pp. 75--136, 2021.

\bibitem{das2020koopman}
S.~Das and D.~Giannakis, ``Koopman spectra in reproducing kernel hilbert
  spaces,'' \emph{Applied and Computational Harmonic Analysis}, vol.~49, no.~2,
  pp. 573--607, 2020.

\bibitem{GoogleLLC}
Google covid-19 community mobility reports.
  \url{https://www.google.com/covid19/mobility/}.

\bibitem{usafacts.org_2021}
Us covid-19 cases and deaths by state.
  \url{https://usafacts.org/visuzalizations/coronavirus-covid-19-spread-map}.

\bibitem{pan2020structure}
S.~Pan and K.~Duraisamy, ``On the structure of time-delay embedding in linear
  models of non-linear dynamical systems,'' \emph{Chaos: An Interdisciplinary
  Journal of Nonlinear Science}, vol.~30, no.~7, p. 073135, 2020.

\bibitem{gao2021early}
X.~Gao, C.~Fan, Y.~Yang, S.~Lee, Q.~Li, M.~Maron, and A.~Mostafavi, ``Early
  indicators of human activity during covid-19 period using digital trace data
  of population activities. front,'' \emph{Built Environ}, vol.~6, p. 607961,
  2021.

\bibitem{hilfiker_josi}
L.~Hilfiker and J.~Josi. epyestim. python package to estimate the time-varying
  effective reproduction number of an epidemic from reported case numbers.
  \url{https://github.com/lo-hfk/epyestim}.

\end{thebibliography}


%
%

\begin{IEEEbiography}[{\includegraphics[width=1in,height=1.25in,clip,keepaspectratio]{\Path 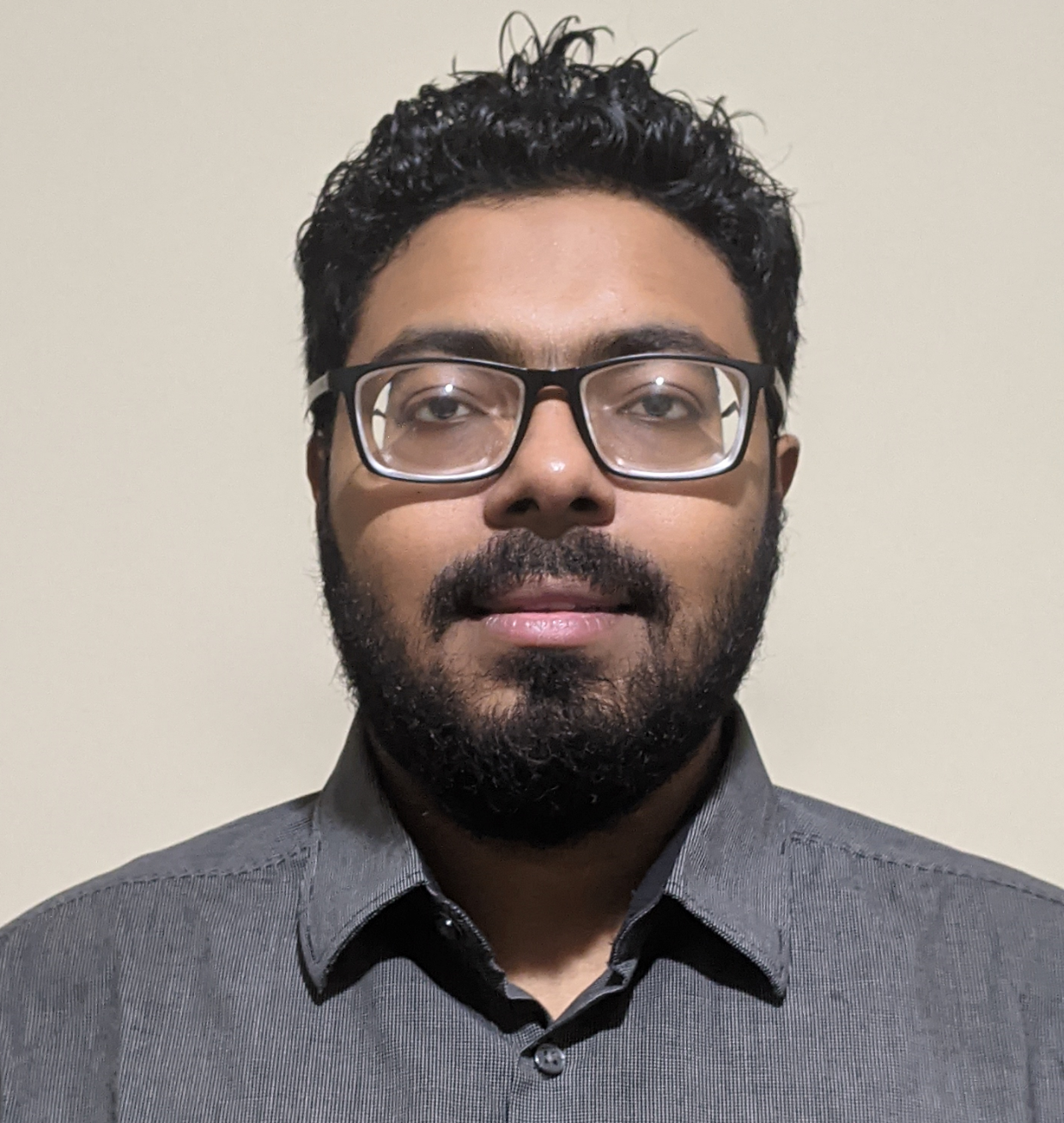}}]%
{Shakib Mustavee}
is a PhD student in the Civil, Environmental, and Construction Engineering Department at the Univerisity  of  Central  Florida. He  is working in the Urban Intelligence and Smart City (URBANITY)Lab under the supervision of Dr. Shaurya Agarwal. He completed his BSc. in Electrical and Electronic Engineering from Bangladesh University of Engineering and Technology in 2017. Currently, his research interests include the application of Dynamic Mode Decomposition and Koopman theory in urban data science.        
\end{IEEEbiography}
\begin{IEEEbiography}[{\includegraphics[width=1in,height=1.25in,clip,keepaspectratio]{\Path 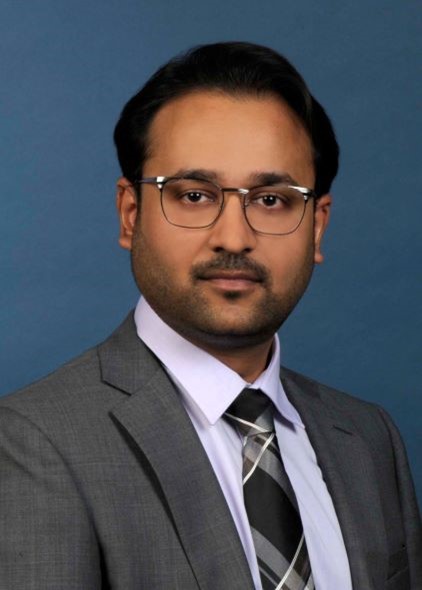}}]%
{Dr. Shaurya Agarwal}
is an Assistant Professor in Civil, Environmental and Construction Engineering Department at University of Central Florida (UCF). He was previously an Assistant Professor in Electrical and Computer Engineering Department at California State University, Los Angeles (CSULA). He completed his Post Doctoral research at New York University and Ph.D. in Electrical Engineering from University of Nevada, Las Vegas (UNLV). His research focuses on interdisciplinary areas of cyber-physical systems, smart and connected transportation, and connected and autonomous vehicles. Passionate about cross-disciplinary research, he integrates control theory, information science, data-driven techniques, and mathematical modeling in his work. He has published over 25 peer-reviewed publications and a book. He is a senior member of IEEE and currently severs as an associate editor of IEEE Transactions on Intelligent Transportation Systems.
\end{IEEEbiography}
%

\begin{IEEEbiography}[{\includegraphics[width=1.0 in,height=1.2 in]{\Path 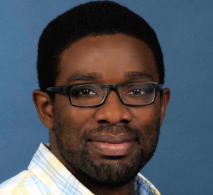}}]
{Dr. Chinwendu Enyioha} is an Assistant Professor in the EECS Department at the University of Central Florida (UCF). Prior to arriving UCF, he was a Postdoctoral Fellow in the EE Department at Harvard University and ME Department at Tufts University, after completing the Ph.D. in Electrical and Systems Engineering at the University of Pennsylvania. His research lies in the areas of decision-making, optimization and resource-aware control of large-scale systems, with applications to multi-robot systems and cyber-physical networks.

\end{IEEEbiography}
\begin{IEEEbiography}[{\includegraphics[width=1.0 in,height = 1.2 in]{\Path 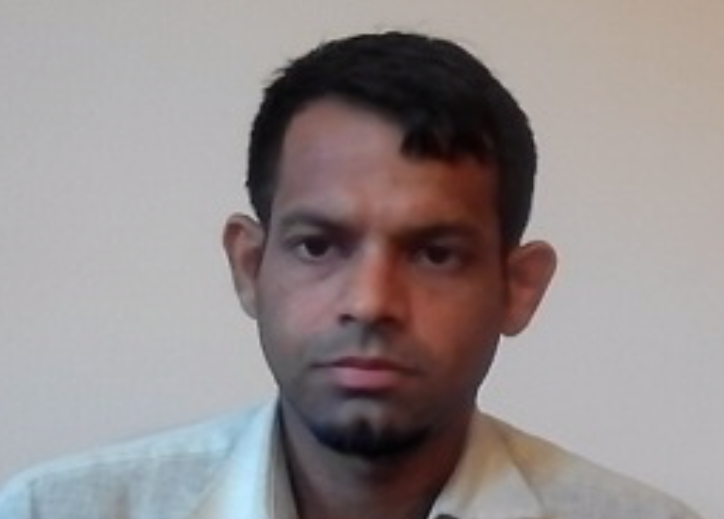}}]
{Dr. Suddhasattwa Das} is a post-doc researcher in the Department of Mathematical Sciences at George Mason University. He completed his Ph.D. in Mathematics at the University of Maryland. He is interested in dynamical systems theory, and its connections to data analysis and signal processing.

\end{IEEEbiography}
\vfill

\end{document}